\documentclass[
letterpaper,
amsmath,
aps, preprint,showpacs,longbibliography
]{revtex4-1}
\usepackage[utf8]{inputenc}  
\usepackage{hyperref}
\usepackage{amsmath}
\usepackage{bm}
\usepackage{amsfonts}
\usepackage{amssymb}
\usepackage{graphicx}
\usepackage{epstopdf}
\usepackage{soul}
\usepackage{xcolor}

\newcommand{\br}{{\bf r}}
\newcommand{\bn}{{\bf n}}

\newcommand{\bu}{{\bf u}}

\newcommand{\bq}{{\bf q}}
\newcommand{\Pe}{{\rm Pe}}
\newcommand{\Tu}{{\rm Tu}}

\newcommand{\bF}{{\bf F}}
\newcommand{\f}{{\bf f}}
\newcommand{\bG}{{\bf G}}
\newcommand{\bp}{{\bf p}}
\newcommand{\bQ}{{\bf Q}}
\newcommand{\bP}{{\bf P}}
\newcommand{\bz}{{\bf z}}

\begin{document}
\title{Swimmer dynamics in externally-driven fluid flows: The role of noise}

\date{\today}

\author{Simon ~A. Berman and Kevin~A. Mitchell}

\affiliation{Department of Physics, University of California, Merced, CA 95344 USA}


\begin{abstract}
We theoretically investigate the effect of random fluctuations on the motion of elongated microswimmers near hydrodynamic transport barriers in externally-driven fluid flows.
Focusing on the two-dimensional hyperbolic flow, we consider the effects of translational and rotational diffusion as well as tumbling, i.e. sudden jumps in the swimmer orientation.
Regardless of whether diffusion or tumbling are the primary source of fluctuations, we find that noise significantly increases the probability that a swimmer crosses one-way barriers in the flow, which block the swimmer from returning to its initial position.
We employ an asymptotic method for calculating the probability density of noisy swimmer trajectories in a given fluid flow, which produces solutions to the time-dependent Fokker-Planck equation in the weak-noise limit. 
This procedure mirrors the semiclassical approximation in quantum mechanics and similarly involves calculating the least-action paths of a Hamiltonian system derived from the swimmer’s Fokker-Planck equation. 
Using the semiclassical technique, we compute (i) the steady-state orientation distribution of swimmers with rotational diffusion and tumbling and (ii) the probability that a diffusive swimmer crosses a one-way barrier.
The semiclassical results compare favorably with Monte Carlo calculations.
\end{abstract}

\maketitle

\section{Introduction}
The advection of self-propelled particles in externally-driven fluid flows presents many surprises when compared to passive advection.
Perhaps the biggest surprise is that the transport efficiency of swimmers does not simply increase as swimming speed increases.
For example, when swimmers are placed in a two-dimensional (2D) oscillating vortex array exhibiting chaotic mixing, faster swimming does not always lead to faster transport \cite{Khurana2011,Khurana2012}.
Even in a steady 2D vortex array, swimmer trapping inside vortices may be enhanced when the particles swim faster, depending on the shape of the particle \cite{Berman2020}.
Similarly, transport efficiency does not simply increase as a swimmer's rotational diffusivity increases, either.
In fact, the opposite occurs in the 2D oscillating vortex array \cite{Torney2007,Khurana2012}.
It is reasonable to expect that the more a swimmer's propulsion direction fluctuates, the smaller its net displacement in a fixed amount of time, and hence the lower the transport efficiency.
Unexpectedly however, the presence of both rotational noise and shear flow can effectively trap swimmers in certain regions, as has been experimentally observed for swimming bacteria \cite{Rusconi2014} and swimming phytoplankton \cite{Barry2015} in a channel flow.
While numerous studies have investigated the spatial distributions of noisy swimmers in a variety of flows \cite{Santamaria2014,Bearon2015,Ezhilan2015,Vennamneni2020,Dehkharghani2019}, a basic understanding of how rotational noise alters swimmer trajectories is lacking.
Our objective in this paper is to develop a theory that quantifies the effect of noise on swimmer dynamics in externally driven fluid flows, especially near transport barriers.

Recently, transport barriers analogous to separatrices---and the related invariant manifolds---of passive advection were identified for self-propelled particles in fluid flows \cite{Berman2021a}.
Perfectly smooth-swimming particles are blocked by so-called swimming invariant manifolds (SwIMs) in position-orientation space.
The SwIMs project to \emph{one-way} barriers, called SwIM edges, to swimmer motion in position space.
Swimmers with orientational noise, on the other hand, can cross SwIM edges, but they are still blocked by one-way barriers known as burning invariant manifolds (BIMs), which were originally introduced as barriers for propagating chemical reaction fronts in fluid flows \cite{Mahoney2012,Mitchell2012}.
By one-way barriers, we mean that swimmers can pass through a BIM or SwIM edge in one direction, but not the other.

\begin{figure}
\centering
\includegraphics[width=\textwidth]{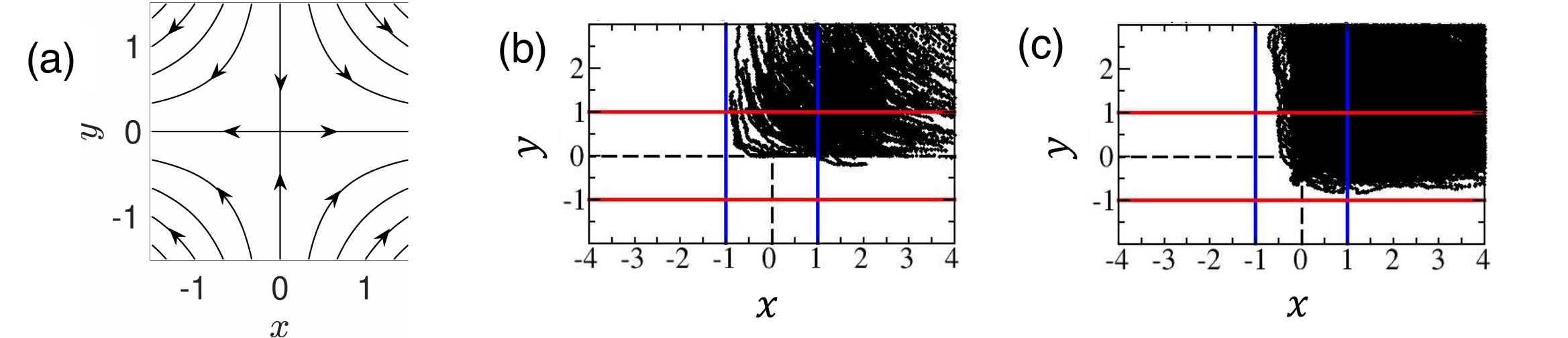}
\caption{Experimental data of swimming \emph{B. subtilis} bacteria trajectories in a microfluidic hyperbolic flow, $\bu = (Bx,-By)$, from Ref.~\cite{Berman2021a}.  $B = 0.44 {\rm s}^{-1}$. (a) Streamlines of the hyperbolic fluid flow. (b) Smooth-swimming bacteria. (c) Run-and-tumble bacteria. The vertical blue lines are SwIM edges blocking inward swimming particles. The horizontal red lines are SwIM edges blocking outward swimming particles. In (b) and (c), each trajectory is plotted in units of $v_0/B$, where $v_0$ is the individual bacterium's measured swimming speed. Every experimentally recorded trajectory is plotted, and the $x \mapsto -x$ and $y \mapsto -y$ symmetries have been used to rectify the trajectories so they all appear to enter from above and exit right.}\label{fig:expt}
\end{figure}
This theory was applied to analyze the experimental trajectories of smooth-swimming and run-and-tumble \emph{Bacillus subtilis} bacteria in a microfluidic cross-channel featuring a hyperbolic fluid flow, illustrated in Fig.~\ref{fig:expt}a.
Whereas the run-and-tumble bacteria exhibit strong rotational noise in the form of sporadic, abrupt changes in swimming direction (Fig.~\ref{fig:expt}c), the smooth-swimming bacteria tend to swim straight in the absence of a flow, with minimal rotational noise (Fig.~\ref{fig:expt}b).
The vertical lines in Figs.~\ref{fig:expt}b and \ref{fig:expt}c are the SwIM edges blocking inward swimming particles, while the horizontal lines are the SwIM edges blocking outward swimming particles.
In the hyperbolic flow, the BIMs coincide exactly with the SwIM edges, and hence these curves are one-way barriers for both the smooth swimming and run-and-tumble bacteria.
Here, all experimentally recorded trajectories are rectified so they appear to enter the flow from above and exit to the right.
Therefore, there can be no trajectories to the left of the SwIM edge $x = -1$, beyond which any trajectory would be swept to the left, as is evident from Figs.~\ref{fig:expt}b and \ref{fig:expt}c.
However, we observe a gap between the left SwIM edge and the measured trajectories of run-and-tumble bacteria in Fig.~\ref{fig:expt}c, compared to the smooth-swimmers in Fig.~\ref{fig:expt}b that can just graze the left SwIM edge before swimming off to the right.
Because the gap represents a depletion of the density of trajectories near the SwIM edge relative to the smooth swimmer case, we refer to it as the depletion effect.
The depletion effect is caused by the orientation fluctuations of the run-and-tumble bacteria.
A run-and-tumble swimmer near this SwIM edge and initially swimming to the right is very likely to tumble and end up crossing the left SwIM edge, forcing it to escape to the left.
At the same time, we observe that the run-and-tumble bacteria swim much closer to the lower SwIM edge than their smooth-swimming counterparts.
This again is due to tumbling.
While smooth swimmers get aligned with the extensional $x$ direction of the flow and thus cannot swim very far  below the line $y=0$, run-and-tumble swimmers can tumble out of alignment and swim towards the lower SwIM edge.
These stark differences between the trajectories of smooth versus run-and-tumble swimmers have motivated the present work.

In this paper, we show how to calculate the probability of particular noisy swimmer trajectories in a given fluid flow, taking the hyperbolic flow as a case study.
Our approach focuses on computing solutions to the time-dependent Fokker-Planck equation of a swimmer (alternately known as a master equation or Smoluchowski equation) in the weak-noise limit \cite{Graham1984,Dykman1996,Gaspard2002,Nolting2016,Gu2020}.
This differs from traditional approaches to the swimmer Fokker-Planck equation, which are focused on the stationary (time-independent) solution and are in the Eulerian frame-of-reference \cite{Rusconi2014,Bearon2015,Ezhilan2015,Vennamneni2020}.
In contrast, we construct a time-dependent swimmer probability density function by following the Lagrangian paths of a swimmer.
This procedure is derived from the weak-noise limit in a manner that is nearly identical to the semiclassical approximation in quantum mechanics \cite{Littlejohn1992}, so we refer to it as the semiclassical approximation to the Fokker-Planck equation.
We use this approach to quantify the depletion of noisy swimmers near a BIM, and compare the results of our semiclassical calculations with Monte Carlo calculations, i.e.\ direct numerical simulations of the swimmer equations of motion.

The paper is organized as follows.
In Sec.~\ref{sec:setup}, we provide background information on the model for swimmer motion employed here and the semiclassical approximation to the Fokker-Planck equation.
In Sec.~\ref{sec:hyp}, we review the dynamics of a deterministic smooth swimmer in the hyperbolic flow, in particular the role of the SwIMs and BIMs.
In Sec.~\ref{sec:orientation}, we compute the position-independent orientation distributions of a swimmer in the hyperbolic flow, obtaining results analogous to the orientation distributions of magnetotactic and viscotactic swimmers in external fields \cite{Waisbord2016,Rupprecht2016,Stehnach2021}, and we apply the semiclassical approximation to calculate the orientation distribution of swimmers with both rotational diffusion and run-and-tumble dynamics.
In Sec.~\ref{sec:depletion}, we compare Monte Carlo and semiclassical calculations of the depletion effect.
Concluding remarks are in Sec.~\ref{sec:conc}.
In the appendix, we present a complete derivation of the semiclassical approximation to the Fokker-Planck equation.

\section{Noisy swimmer dynamics}\label{sec:setup}
We consider the motion of an ellipsoidal swimmer in two dimensions, with position $\br = (x,y)$ and orientation $\hat{\bn} = (\cos\theta,\sin\theta)$.
The stochastic differential equations describing a noisy swimmer in a fluid flow ${\bu}( \br)$ are \cite{Torney2007,Khurana2012}
\begin{subequations}\label{eq:sde}
\begin{align}\label{eq:sde_r}
& {\rm d} \br = \left[\bu(\br) + v_0 \hat{\bn} \right] {\rm d} t + \sqrt{2 D_T} \,{\rm d} w_\br,\\ \label{eq:sde_th}
& {\rm d} \theta = \left[ \frac{\omega(\br)}{2} + \alpha \hat{\bn}_\perp \cdot {\mathsf E}(\br) \hat{\bn}  \right] {\rm d} t + \sqrt{2 D_R} \, {\rm d} w_\theta +{\rm d} L(\nu),
\end{align}
\end{subequations}
where $\omega = \hat{\bf z} \cdot (\nabla \times \bu)$ is the vorticity, $\hat{\bn}_\perp = (-\sin \theta,\cos \theta)$, and ${\mathsf E} = (\nabla \bu + \nabla \bu^{\rm T})/2$ is the symmetric rate-of-strain tensor.
The shape parameter $\alpha$ equals  $(a^2-1)/(a^2+1)$, where $a$ is the aspect ratio of the ellipse; $\alpha$  varies from $-1$ to $1$, where $\alpha = 0$ is a circle,  and $|\alpha|=1$ is an infinitely thin rod.
Positive (negative) values of $\alpha$ correspond to swimming parallel (perpendicular) to the major axis. 
Each equation of \eqref{eq:sde} contains a deterministic drift term, proportional to ${\rm d}t$, and noise terms.
In Eq.~\eqref{eq:sde_r}, the noise terms are the independent Wiener processes $w_\br = (w_x,w_y)$ and account for translational diffusion with diffusivity $D_T$.
Note that for certain swimmers, the strength of the translational diffusivity along the particle's major axis may differ from the translational diffusivity along the minor axis, and their may be additional correlations between translational and rotational noise \cite{Thiffeault2021}.
We ignore these issues here for simplicity.


Equation \eqref{eq:sde_th}, on the other hand, contains two stochastic terms describing fluctuations in the swimming direction.
We distinguish between two types of rotational noise: rotational diffusion and tumbling.
Rotational diffusion refers to continual random perturbations in the swimmer orientation, such that in the absence of a flow, the orientation $\theta$ would exhibit free diffusion (given by the Wiener process $w_\theta$) with a rotational diffusivity $D_R$.
This can arise due to random fluctuations in the propulsion force of the swimmer \cite{Hyon2012,Thiffeault2021} or from the thermal fluctuations of the surrounding fluid.
In the former case, the noise mechanism would lead to correlations between translational and rotational diffusion, but here we neglect those for simplicity.
Tumbling refers to the sudden resetting of $\theta$ to a random orientation, independent of its present value, which occurs sporadically as a Poisson process $L(\nu)$ with frequency $\nu$.
This kind of sudden, random reorientation is seen in swimming bacteria in the ``run-and-tumble" mode of swimming.
In practice, the distribution of new orientations may depend on the previous value, as is the case for wild-type strains of the swimming bacteria \emph{E. coli} \cite{Berg1972}, but we neglect this here for simplicity.
Hence, the new angle after a tumble is uniformly and randomly distributed between $0$ and $2\pi$.

Our goal in this paper is to estimate the probability of various swimmer trajectories of Eq.~\eqref{eq:sde}.
This can certainly be accomplished by Monte Carlo simulations, i.e.\ direct numerical simulations of Eq.~\eqref{eq:sde}, but we also develop an analytical approach to calculating such probabilities, which is less computationally costly and provides deeper theoretical insight into the swimmer dynamics.
To study the probability of swimmer trajectories, we investigate the Fokker-Planck equation for the probability density of the particle $P({\bf r},\theta,t)$ \cite{Solon2015},
\begin{equation}\label{eq:mastereq}
\frac{\partial P}{\partial t} = -\nabla \cdot (\bF P) + \frac{\varepsilon}{2}\left[ \gamma\left(\frac{\partial^2 P}{\partial x^2} + \frac{\partial^2 P}{\partial y^2} \right) + \frac{\partial^2 P}{\partial \theta^2} \right]+ \lambda \left[ -P + \frac{1}{2\pi} \int_0^{2\pi} P(\br,\theta',t){\rm d} \theta' \right],
\end{equation}
where we have abbreviated the deterministic drift terms from Eq.~\eqref{eq:sde} as $\bF$, with
\begin{equation}
\bF = \left(u_x+ v_0 \cos \theta,\,\, u_y + v_0 \sin \theta,\,\, \frac{\omega}{2} + \alpha \hat{\bf n}_\perp \cdot {\mathsf E} \hat{\bf n} \right).
\end{equation}
Here, $\nabla = (\partial/\partial x, \partial/ \partial y, \partial/ \partial \theta)$.
We have non-dimensionalized the coordinates using a length scale $\mathcal{L}$ and velocity scale $\mathcal{U}$,
so that
\begin{equation}
\varepsilon =  \frac{2D_R\mathcal{L}}{\mathcal{U}}
\end{equation}
is the strength of rotational diffusion, $\gamma = D_T/(\mathcal{L}^2 D_R)$ is the ratio of translational diffusion to rotational diffusion (usually $\gamma  < 1$), and  $\lambda =\nu \mathcal{L}/\mathcal{U}$ is the non-dimensional tumbling rate.
Note that the rotational P\'{e}clet number is $\Pe = 2/\varepsilon$ \cite{Ezhilan2015}.
The first two terms on the right-hand side of Eq.~\eqref{eq:mastereq} are the usual drift and diffusion terms.
The third term proportional to $\lambda$ accounts for tumbling, with the first term in brackets describing the loss of probability due to tumbling out of the present angle, and the second term describing the gain of probability from the swimmers at all other angles that have tumbled into the present angle.
Equation \eqref{eq:mastereq} is difficult to attack in general, so we focus on special cases where exact or approximate analytical (or semi-analytical) solutions may be found.

Of particular interest is the  $\lambda = 0$ case, describing non-tumbling swimmers or, alternatively, the evolution of the probability density in between tumble events.
In this case we seek asymptotic solutions in the weak diffusion ($\varepsilon \ll 1$) limit, which have the WKB form
\begin{equation}\label{eq:WKB}
P(\br,\theta,t) \approx A(\br,\theta,t) \exp\left[ -\frac{W(\br,\theta,t)}{\varepsilon} \right].
\end{equation}
Equation \eqref{eq:WKB} has the same form as the semiclassical approximation to the wave function in quantum mechanics, and hence we refer to it as the semiclassical approximation.
Substituting this approximation into Eq.~\eqref{eq:mastereq} with $\lambda = 0$ leads to a Hamilton-Jacobi equation for $W$ and a related transport equation for $A$.

Here, we briefly describe the mathematical theory of approximation \eqref{eq:WKB}, while a detailed derivation and discussion are contained in Appendix \ref{sec:app}.
The solution $W$ to the Hamilton-Jacobi equation \eqref{eq:HJ} is related to the classical action accumulated along the trajectories of a particular Hamiltonian system associated with Eq.~\eqref{eq:mastereq}.
The Hamiltonian is given by 
\begin{equation}\label{eq:Hamiltonian}
H(\bq,\bp) = \frac{1}{2}\left[ \gamma(p_x^2 + p_y^2) + p_\theta^2 \right] + \bp \cdot \bF(\bq),
\end{equation}
where $\bq = (x,y,\theta)$ and $\bp = (p_x,p_y,p_\theta)$.
Equation \eqref{eq:Hamiltonian} follows from the Hamiltonian \eqref{eq:ham} for a general Fokker-Planck equation \eqref{eq:fp}, of which Eq.~\eqref{eq:mastereq} (with $\lambda = 0$) is a special case.
The function $W$ is equivalent to the Onsager-Machlup-Freidlin-Wentzell action function \cite{Onsager1953a,Freidlin2012,Gaspard2002} which arises in nonequilibrium statistical mechanics \cite{Graham1984,Dykman1996} and rare event modeling \cite{Forgoston2018}.
At each point $(\br,\theta)$, the action $W(\br,\theta,t)$ can be expressed as an integral along the trajectory of the Hamiltonian system with Hamiltonian \eqref{eq:Hamiltonian} that arrives at that point at time $t$.
This makes Eq.~\eqref{eq:WKB} a Lagrangian, as opposed to Eulerian, description of the probability density.
The trajectories associated with the minima of $W$, i.e. the minimum-action paths, correspond to the most likely trajectories of a noisy swimmer, because at lowest order in $\varepsilon$, the probability density \eqref{eq:WKB} is peaked at these points.
Hence, finding the minimum-action paths is the main focus of most works involving the Onsager-Machlup-Freidlin-Wentzell action function, including recent work on escape paths of active particles in potential wells \cite{Gu2020}.
In contrast, we consider all possible paths, in order to get a global approximation to the probability density \eqref{eq:WKB}.

Throughout the paper, we focus on the hyperbolic flow $\bu = (Bx,-By)$.
Therefore, Eq.~\eqref{eq:sde} becomes
\begin{subequations}\label{eq:hypflow}
\begin{align} \label{eq:xdot}
& {\rm d} x = (x+ \cos\theta) {\rm d} t + \sqrt{\varepsilon \gamma } {\rm d}w_x, \\ \label{eq:ydot}
& {\rm d} y = (-y + \sin \theta) {\rm d} t + \sqrt{\varepsilon \gamma} {\rm d} w_y  \\ \label{eq:thdot}
& {\rm d}\theta = -\alpha \sin(2 \theta) {\rm d} t + \sqrt{\varepsilon} {\rm d} w_\theta + {\rm d} L(\lambda).
\end{align}
\end{subequations}
We have taken the velocity scale $\mathcal{U} = v_0$ and the length scale $\mathcal{L} = v_0/B$ in the non-dimensional equation \eqref{eq:hypflow}.
The typical values of $\varepsilon$, $\gamma$, $\lambda$, and $\alpha$ depend on the system being considered.
In the hyperbolic flow experiments leading to Fig.~\ref{fig:expt}, $B = 0.44\,\,{\rm s}^{-1}$ \cite{Berman2021a}.
The measured rotational diffusivity for several species of swimming phytoplankton is $D_R = 0.15$--$0.27\,\,{\rm rad}^2 / {\rm s}$ \cite{Barry2015}, which in the hyperbolic flow would yield $\varepsilon = 0.68$--$1.2$.
For wild-type \emph{E.~coli}, which exhibit run-and-tumble behavior, the rotational diffusivity during runs is $D_R = 0.06\,\,{\rm rad}^2/{\rm s}$ \cite{Locsei2009} and the tumbling frequency is approximately $\nu = 1\,\,{\rm s}^{-1}$ \cite{Berg1972}, which in the hyperbolic flow would yield $\varepsilon = 0.27$ and $\lambda = 2.3$.
Assuming that the translational diffusion for wild-type \emph{E.~coli} is due only to thermal fluctuations so that $D_T = 0.2\,\,\mu {\rm m}^2/ {\rm s}$ \cite{Bearon2015} and a swimming speed $v_0 = 14\,\,\mu {\rm m}/s$ \cite{Berg1972}, the translational-to-rotational diffusion ratio would be $\gamma = 0.003$.
Note that in the hyperbolic flow, $\varepsilon$ and $\lambda$ can always be made smaller and $\gamma$ larger by increasing $B$, the flow strength parameter.
We take $\alpha = 1$ in all numerical computations, corresponding to the elongated shape of swimming bacteria like \emph{E.~coli} and \emph{B.~subtilis}.
Numerical solutions of Eq.~\eqref{eq:hypflow} are obtained using the Euler-Maruyama method.
Before investigating the dynamics of noisy swimmers in the hyperbolic flow, we study the deterministic dynamics of Eq.~\eqref{eq:hypflow} with $\varepsilon = 0$ and $\lambda = 0$.

\section{Deterministic dynamics in the hyperbolic flow}\label{sec:hyp}
\begin{figure}
\centering
\includegraphics[width=\textwidth]{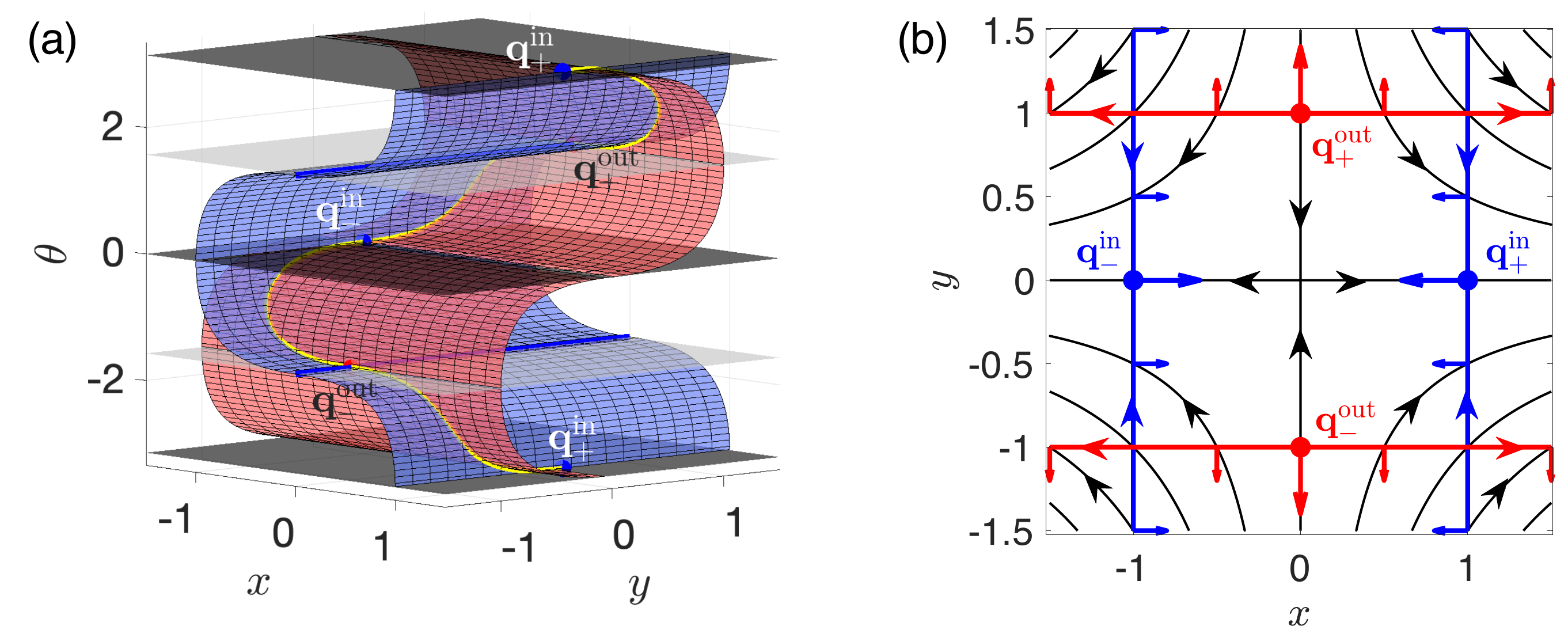}
\caption{Phase space structure of a swimmer in the hyperbolic flow, for $\alpha = 1$. (a) Swimming fixed points (SFPs) and their invariant manifolds (SwIMs). The red surfaces are the unstable SwIMs of $\bq^{\rm out}_\pm$, and the blue surfaces are the stable SwIMs of $\bq^{\rm in}_\pm$. The dark blue lines are the 1D stable SwIMs of $\bq^{\rm out}_\pm$. The dark and light grey planes are constant-$\theta$ invariant surfaces which are displayed for visualization purposes. (b) Projection of swimming fixed points and SwIMs into the $xy$ plane. The black curves are the streamlines of the hyperbolic flow. The red (blue) lines are the unstable (stable) SwIM edges. The small arrows perpendicular to the SwIM edges point in the swimming direction.}\label{fig:swims3Dxy}
\end{figure}
The deterministic dynamics of Eq.~\eqref{eq:hypflow} is best understood through the system's fixed points and invariant manifolds, previously studied in Ref.~\cite{Berman2021a}.
The system possesses four fixed points, which we refer to as swimming fixed points (SFPs) to distinguish them from the passive fixed points of the fluid flow.
Denoting the phase-space coordinate $\bq = (x,y,\theta)$, the fixed points are
\begin{equation}
\bq_{\pm}^{\rm out} = \left(0,\pm 1, \pm \frac{\pi}{2}\right), \quad \bq^{\rm in}_+ = (1,0,\pi), \quad \bq^{\rm in}_{- }= (-1,0,0),
\end{equation}
illustrated in Fig.~\ref{fig:swims3Dxy}.
Each of the SFPs is a saddle.
When $\alpha > 0$, and in particular when $\alpha = 1$, the $\bq_\pm^{\rm out}$ fixed points have stable-unstable-unstable (SUU) linear stability, and the $\bq_\pm^{\rm in}$ fixed points have stable-stable-unstable (SSU) linear stability.
Hence, the $\bq_\pm^{\rm out}$ SFPs possess 2D unstable manifolds, while the $\bq_\pm^{\rm in}$ fixed points possess 2D stable manifolds.
We refer to these 2D manifolds as \emph{swimming invariant manifolds} (SwIMs), to distinguish them from the invariant manifolds of passive advection \cite{Berman2021a}.

Taken as a whole, the stable and unstable SwIMs consist of two interlocking S-shaped sheets, plotted in Fig.~\ref{fig:swims3Dxy}a.
The stable SwIMs  (the blue surface) attached to $\bq_+^{\rm in}$ and $\bq_-^{\rm in}$ share common boundaries along the lines $\{(x,y,\theta) \,\,|\,\, x = 0,\,\,\theta = \pm \pi\}$ which are the 1D stable manifolds of $\bq_\pm^{\rm out}$ (dark blue lines).
Hence, the union of the 2D stable SwIMs with the 1D stable manifolds of  $\bq_\pm^{\rm out}$ is a surface (blue surface in Fig.~\ref{fig:swims3Dxy}a) which separates phase space into two pieces.
By symmetry, the same geometric shape can be constructed by taking the union of the unstable SwIMs of $\bq_\pm^{\rm out}$ with the 1D unstable manifolds of $\bq_\pm^{\rm in}$, leading to the red surface in Fig.~\ref{fig:swims3Dxy}a.
The shape of the stable SwIMs is independent of the $y$ coordinate and, similarly, the shape of the unstable SwIMs is independent of the $x$ coordinate.
This occurs because in the hyperbolic flow, the $x\theta$ equations Eq.~\eqref{eq:xdot} and \eqref{eq:thdot} are decoupled from $y$ and similarly the $y\theta$ equations Eq.~\eqref{eq:ydot} and \eqref{eq:thdot} are decoupled from $x$.
The stable and unstable SwIMs intersect along heteroclinic orbits going from one fixed point to another, indicated by the yellow curves in Fig.~\ref{fig:swims3Dxy}a.

\begin{figure}
\centering
\includegraphics[width=\textwidth]{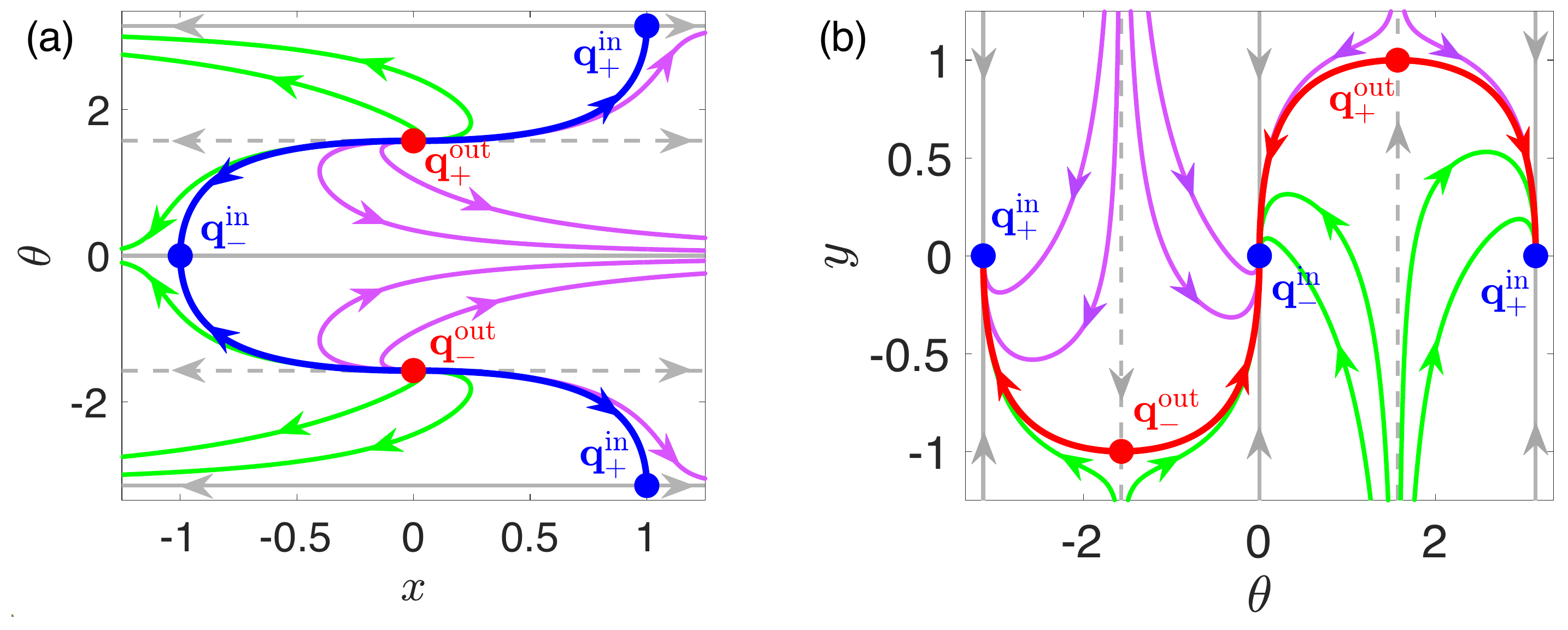}
\caption{Phase portraits of swimmer dynamics in the hyperbolic flow, for $\alpha = 1$. (a) $x\theta$ cross-section of the dynamics. The blue curve is the cross-section of the stable SwIM of $\bq_\pm^{\rm in}$. (b) $y\theta$ cross-section of the dynamics. The red curve is the cross-section of the unstable SwIM of $\bq_\pm^{\rm out}$. In both panels, the solid and dotted grey lines are cross-sections of the stable and unstable constant-$\theta$ invariant surfaces, respectively.}\label{fig:swims2D}
\end{figure}

Cross-sections of the SwIMs are shown in Fig.~\ref{fig:swims2D}, along with the phase portraits of the $x\theta$ dynamics and the $y\theta$ dynamics.
Figure \ref{fig:swims2D}a shows that swimmers on the left of the stable SwIM ultimately exit the flow to the left, while swimmers on the right ultimately exit right.
Similarly, the unstable SwIM (Fig.~\ref{fig:swims2D}b) separates swimmers that entered the flow from the top from those which entered from the bottom.
The SwIMs are therefore transport barriers to swimmers in the hyperbolic flow, because they carve out the $xy\theta$ phase space into distinct, qualitatively different families of trajectories.
Importantly, these barriers are nonporous in phase space, meaning no swimmer trajectory can cross them (in the absence of noise).

On the other hand, in position space, the SwIMs project to one-way barriers, allowing swimmers to cross in one direction but not the other.
Figure \ref{fig:swims3Dxy}b shows the singularities of the projection of the SwIMs into the $xy$ plane---that is, the folds of the S-shapes which bound the projection of the 2D surfaces into the plane.
We refer to these curves as SwIM edges \cite{Berman2021a}.
The stable SwIM edges at $x = \pm 1$ (blue curves in Fig.~\ref{fig:swims3Dxy}b) block inward swimming particles, while allowing outward swimming particles through.
To see this, note that for $x = -1$, $\dot{x} \leq 0$ for all $\theta$, and for $x = 1$, $\dot{x} \geq 0$ for all $\theta$, as shown in Fig.~\ref{fig:swims2D}a.
Along the stable SwIM edges, the outward fluid flow overpowers the swimmers and they are swept away from the center of the flow.
Similarly, the unstable SwIM edges (red curves in Fig.~\ref{fig:swims3Dxy}b) block outward swimming particles, while inward swimming particles can pass through them.
Here, for $y = 1$, $\dot{y} \leq 0$, and for $y = -1$, $\dot{y} \geq 0$.
On the unstable SwIM edges, it is the inward flow which overpowers the swimmers and pushes them towards the center of the flow.

The SwIM edges in the hyperbolic flow coincide exactly with the BIMs---the 1D invariant manifolds of the SFPs when $\alpha = -1$ \cite{Berman2021a,Mahoney2015}.
This is important because SwIM edges are only guaranteed to be one-way barriers for purely deterministic swimmers.
BIMs, on the other hand, have stronger barrier properties, in that they are also one-way barriers for swimmers with rotational diffusion or run-and-tumble dynamics in the limit of negligible translational diffusion \cite{Berman2021a}.
Thus, in the hyperbolic flow, the SwIM edges also act as one-way barriers for swimmers with rotational noise.
This explains why the run-and-tumble bacteria in the hyperbolic flow experiment remain bounded by the unstable SwIM edge at $y = -1$ in Fig.~\ref{fig:expt}b.
Similarly, the stable SwIM edges act as points of no return for all swimmers.
Once a swimmer swims over a stable SwIM edge, it is unable to swim back to the center of the flow.
This is the origin of the depletion effect we observe when comparing the smooth swimming bacteria data (Fig.~\ref{fig:expt}a) to the run-and-tumble data (Fig.~\ref{fig:expt}b).
The orientation fluctuations of tumbling bacteria make it very likely that a bacterium near the SwIM edge at $x = -1$, for example, swims across it, precluding the possibility that it subsequently exits the flow to the right.
Hence, we expect to observe much fewer trajectories of run-and-tumble swimmers initially near the left SwIM edge and exiting right relative to smooth swimmers, consistent with the experimental data.

\section{Steady-state orientation distributions in the hyperbolic flow}\label{sec:orientation}
Because the $\dot{\theta}$ equation \eqref{eq:thdot} is independent of $x$ and $y$, we begin by looking at the effect of noise on the orientation dynamics alone in the hyperbolic flow.
The Fokker-Planck equation for the probability density $P(\theta,t)$ restricted to the orientation degree-of-freedom is 
\begin{equation}\label{eq:fokk_theta}
\frac{\partial P}{\partial t} = -\frac{\partial}{\partial \theta} [ -\alpha \sin (2\theta) P] + \frac{\varepsilon}{2} \frac{\partial^2 P}{\partial \theta^2} + \lambda \left( -P + \frac{1}{2\pi}\right).
\end{equation}
We focus on the stationary distributions of the $\theta$ variable, which are the stationary solutions ($\partial P/\partial t = 0$) of Eq.~\eqref{eq:fokk_theta}.
We first treat the two limiting cases (i) no tumbling ($\lambda = 0$), and (ii) no rotational diffusion ($\varepsilon = 0$), before proceeding to the case where there is both tumbling and rotational diffusion.
Note that the orientation dynamics of a noisy swimmer in the hyperbolic flow is very similar to the orientation dynamics of  swimmers in other types external fields, such as magnetotactic swimmers in external magnetic fields \cite{Rupprecht2016,Waisbord2016} and swimmers in viscocity gradients \cite{Stehnach2021}.
The main difference here compared to the preceding examples, aside from the source of the torque on the swimmer, is that Eq.~\eqref{eq:fokk_theta} is invariant under the symmetry $\theta \mapsto \theta+\pi$.

\subsection{Orientation dynamics with rotational diffusion only ($\lambda = 0$)}
\begin{figure}[t!]
\centering
\includegraphics[width=0.8\textwidth]{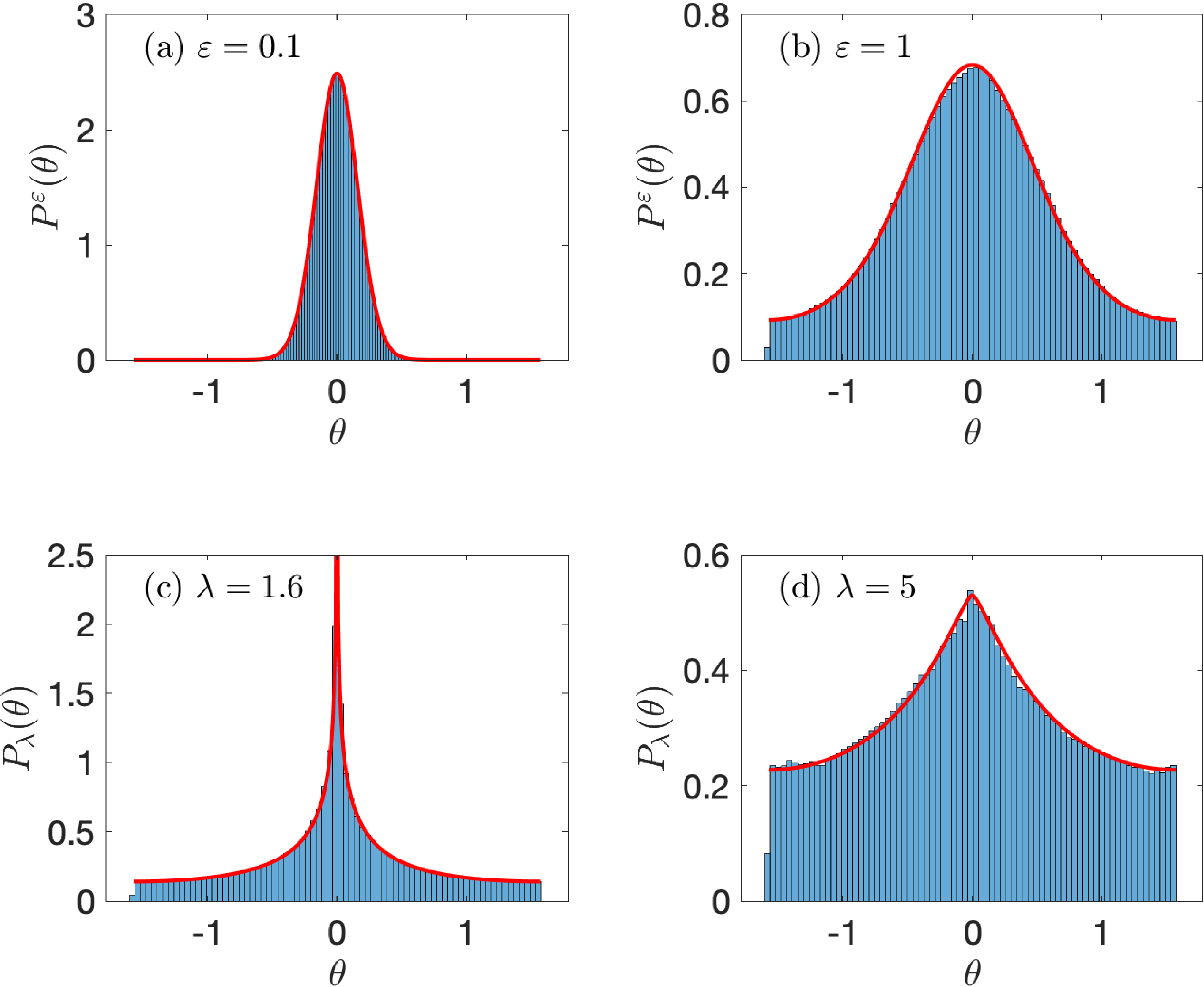}
\caption{Stationary $\theta$ distributions with $\alpha = 1$, with no tumbling (a), (b), and no rotational diffusion (c), (d). Histograms are Monte Carlo simulations of Eq.~\eqref{eq:thdot}, and red curves are the theoretically predicted distributions given by Eq.~\eqref{eq:theta_dist} for the no-tumbling case and Eq.~\eqref{eq:theta_dist_rt} for the no-diffusion case. Distributions are plotted in the range $\theta \in (-\pi/2,\pi/2)$. (a) $\lambda = 0$, $\varepsilon = 0.1$. (b) $\lambda = 0$, $\varepsilon = 1$. (c) $\varepsilon = 0$, $\lambda = 1.6$. (d) $\varepsilon=0$, $\lambda = 5$.}\label{fig:theta_rt_dist}
\end{figure}

Without tumbling, the $\theta$ dynamics is governed by
\begin{equation}\label{eq:theta_noise}
{\rm d} \theta = -\alpha \sin( 2 \theta) {\rm d} t + \sqrt{\varepsilon} {\rm d}w_\theta.
\end{equation}
The deterministic part of the equation has the form of the gradient of a potential $V(\theta)$, meaning we can write ${\rm d} \theta = -(\partial V/\partial \theta){\rm d} t + \sqrt{\varepsilon} {\rm d}w_\theta$, with $V(\theta) = -\alpha \cos (2\theta)/2$.
Hence, the dynamics is equivalent to that of an overdamped particle in the potential $V(\theta)$ with noisy driving.
In this case, the long-time probability distribution of $\theta$ evolves towards a stationary state which is peaked at the potential wells at $\theta = 0$ and $\theta = \pi$ (for $\alpha > 0$).
This probability distribution $P^\varepsilon(\theta)$ can be found by solving for the stationary state of Eq.~\eqref{eq:fokk_theta} with $\lambda = 0$.
For gradient systems, the solution is simply $P^\varepsilon(\theta) \propto \exp[-2 V(\theta)/\varepsilon]$, which is simple to verify, and hence we have
\begin{equation}
P^\varepsilon(\theta) \propto \exp\left[\frac{\alpha}{\varepsilon}\cos 2\theta\right].
\end{equation}
Clearly, the distribution depends on a single dimensionless parameter, 
\begin{equation}
\frac{\alpha}{\varepsilon} = \frac{A \alpha}{2 D_R},
\end{equation}
which is the ratio of the rate of alignment with the extensional direction of the flow, $A\alpha$, to the intensity of the noise.
Normalizing the probability distribution, we obtain
\begin{equation}\label{eq:theta_dist}
P^\varepsilon(\theta) = \left[2 \pi I_0 \left(\frac{\alpha}{\varepsilon}\right) \right]^{-1} \exp\left[\frac{\alpha}{\varepsilon} \cos 2\theta\right],
\end{equation}
where $I_0(x)$ is a modified Bessel function of the first kind.
The stationary distribution \eqref{eq:theta_dist} is invariant under the shift symmetry $\theta \mapsto \theta + \pi$, as is the underlying stochastic process \eqref{eq:theta_noise}.
Equation \eqref{eq:theta_dist} is plotted in Figs.~\ref{fig:theta_rt_dist}a and \ref{fig:theta_rt_dist}b, along with histograms from Monte Carlo simulations of Eq.~\eqref{eq:theta_noise}.
In Fig.~\ref{fig:theta_rt_dist}, we map the Monte Carlo data onto the interval $\theta \in (-\pi/2,\pi/2)$ using symmetry and only plot Eq.~\eqref{eq:theta_dist} in this range.
Similar results were previously obtained for magnetotactic swimmers in an external magnetic field with rotational diffusion \cite{Rupprecht2016}.

\subsection{Orientation dynamics with tumbling only ($\varepsilon = 0$)}\label{sec:tumbling}
Here we consider the case of the stationary $\theta$ distribution under tumbling only.
Every time a swimmer tumbles, its orientation is drawn from the uniform distribution.
If it tumbles at time $\tau = 0$, then until the next tumble, its probability density $P(\theta,\tau)$ evolves according to the Liouville equation
\begin{equation}\label{eq:theta_liou}
\frac{\partial P}{\partial \tau} = \frac{\partial}{\partial \theta}\left[\alpha \sin (2\theta)P\right],
\end{equation}
with the initial condition $P(\theta,0) = P_0(\theta) = 1/2\pi$.
Intuitively, we thus expect that the steady-state distribution under tumbling only, $P_\lambda(\theta)$, should consist of the superposition of probability distributions $P(\theta,\tau)$ describing the relaxation of $\theta$ in between tumbles, weighted by the probability $\lambda e^{-\lambda \tau} {\rm d} \tau$ that the last tumble occurred a time $\tau$ in the past.
In other words, the stationary probability distribution must be \cite{Evans2020}
\begin{equation}\label{eq:rt_dist}
P_\lambda(\theta) = \lambda \int_0^\infty P(\theta,\tau) e^{-\lambda \tau} {\rm d} \tau.
\end{equation}
It is straightforward to verify that Eq.~\eqref{eq:rt_dist} is a stationary solution of Eq.~\eqref{eq:fokk_theta} with $\varepsilon = 0$.

An explicit solution to Eq.~\eqref{eq:theta_liou} can be obtained using the method of characteristics, because the $\theta$ equation of motion \eqref{eq:theta_noise} in the absence of noise ($\varepsilon = 0$) can be solved analytically.
The expression for the deterministic trajectory $\theta^*(t)$ is
\begin{equation}\label{eq:theta_t_exact}
\theta^*(t) = \tan^{-1}\left(e^{-2 \alpha t} \tan \theta_0  \right).
\end{equation}
where $\theta_0 = \theta(0)$ is the initial condition, and here it is assumed $\theta_0 \in [-\pi/2,\pi/2]$.
This condition arises due to the use of the $\tan^{-1}$ function; when $\theta_0$ is outside this range, this solution needs to be shifted either up or down by $\pi$, depending on $\theta_0$.
Then, the solution to Eq.~\eqref{eq:theta_liou} is
\begin{equation}\label{eq:theta_dist_t}
P(\theta,\tau) = P_0\left(\tan^{-1}\left(e^{2  \alpha \tau} \tan \theta \right) \right)\frac{e^{2\alpha\tau}}{\cos^2 \theta +e^{4 \alpha \tau}\sin^2\theta },
\end{equation}
where $P_0(\theta) = P(\theta,0)$ is an arbitrary initial orientation distribution (see Appendix \ref{sec:liou_solu} for the derivation).
Again, this form of the solution is valid for $\theta \in [-\pi/2,\pi/2]$, and a shift by $\pi$ in the argument of $P_0$ in Eq.~\eqref{eq:theta_dist_t} adapts the solution to the excluded range of $\theta$.

Next, we obtain the stationary $\theta$ distribution $P_\lambda(\theta)$ under tumbling with rate $\lambda$ by substituting Eq.~\eqref{eq:theta_dist_t} into Eq.~\eqref{eq:rt_dist}, with $P_0 = 1/2\pi$.
Rescaling the time in Eq.~\eqref{eq:rt_dist} by the tumbling rate $\lambda$, we obtain a complicated integral that depends on a single dimensionless parameter that we call the tumbling number $\Tu$,
\begin{equation}\label{eq:Tu}
\Tu = \frac{\lambda}{2 \alpha} = \frac{\nu}{2 A \alpha}.
\end{equation}
This is essentially the ratio of the tumbling rate to the relaxation rate of a swimmer's orientation to its equilibrium (parallel to the extensional $x$-direction) in the hyperbolic flow.
Note, the latter relaxation rate is distinct from the relaxation rate of the orientation distribution of a swimmer with rotational diffusion to the stationary state given by Eq.~\eqref{eq:theta_dist}.
It is conceivable that this is the more relevant time scale for defining $\Tu$ in the case where we have both tumbling and rotational diffusion.
The stationary distribution $P_\lambda$ can be shown to be equal to
\begin{equation}\label{eq:theta_dist_rt}
P_\lambda(\theta) = \frac{\Tu}{2 \pi} \,\frac{_2 F_1(1,(1 +\Tu)/2;(3+\Tu)/2;-\cot^2 \theta)}{(1 + \Tu)\sin^2\theta },
\end{equation}
where $_2F_1(a,b;c;z)$ is the ordinary hypergeometric function.
Like $P^\varepsilon(\theta)$ (Eq.~\eqref{eq:theta_dist}), Eq.~\eqref{eq:theta_dist_rt} is invariant under the shift symmetry $\theta \mapsto \theta + \pi$.
To be sure, the hypergeometric function makes the expression \eqref{eq:theta_dist_rt} of the tumbling swimmer's stationary distribution more opaque than its counterpart for rotational diffusion.

In Figs.~\ref{fig:theta_rt_dist}c and \ref{fig:theta_rt_dist}d, we plot Eq.~\eqref{eq:theta_dist_rt}, superimposed over histograms from numerical simulations of tumbling swimmers in the hyperbolic flow, to obtain a basic intuition for how the distribution depends on the parameters.
We see an excellent agreement between the theory and simulations.
For sufficiently small tumbling rates such that $\Tu \leq 1$ in Eq.~\eqref{eq:Tu} (Fig.~\ref{fig:theta_rt_dist}c), it can be shown that $P_\lambda$ is singular at the orientation equilibrium $\theta = 0$ and relatively flat for all other orientations.
On the other hand, for $\Tu > 1$ (Fig.~\ref{fig:theta_rt_dist}d), the peak at the equilibrium becomes finite, and the difference between the probabilities near the stable and unstable equilbria becomes more modest.
Again, our results mirror those obtained for magnetotactic run-and-tumble bacteria in Ref.~\cite{Rupprecht2016}.

\subsection{Orientation distribution with rotational diffusion and tumbling}
Having treated the limiting cases of one type of noise versus another, we now seek the stationary $\theta$ distribution of a run-and-tumble swimmer with rotational diffusion, which we denote $P^\varepsilon_\lambda(\theta)$.
This requires slightly modifying the approach of Sec.~\ref{sec:tumbling}, where the stationary distribution is the weighted time average of the probability distributions describing relaxation to equilibrium, $P(\theta,\tau)$ [see Eq.~\eqref{eq:rt_dist}].
Namely, when we have both tumbling and rotational diffusion, we need to obtain  $P(\theta,\tau)$ by solving the time-dependent Fokker-Planck equation 
\begin{equation}\label{eq:fokk_theta_no_tumble}
\frac{\partial P}{\partial \tau} = \frac{\partial}{\partial \theta} [ \alpha \sin (2\theta) P] + \frac{\varepsilon}{2} \frac{\partial^2 P}{\partial \theta^2},
\end{equation}
with initial condition $P_0(\theta) = 1/2\pi$, instead of the Liouville equation \eqref{eq:theta_liou}.
An exact analytical solution of Eq.~\eqref{eq:fokk_theta_no_tumble} is unavailable, so we resort to a short-time, small $\varepsilon$ asymptotic approximation based on the semiclassical techniques detailed in Sec.~\ref{sec:gauss_approx}.

\subsubsection{Approximation of the Fokker-Planck propagator for small $\varepsilon$}
$P(\theta,\tau)$ is the probability density of reaching $\theta$ at time $\tau$ under Eq.~\eqref{eq:theta_noise} with a random initial condition drawn from a uniform distribution.
We approximate $P(\theta,\tau)$ by making use of the semiclassical approximation to the Fokker-Planck equation \eqref{eq:fokk_theta_no_tumble}.
We note that this distribution can be expressed as an integral over the propagator $K(\theta,\theta_0,\tau)$, which is the probability distribution of reaching $\theta$ in time $\tau$ from a fixed initial condition $\theta_0$, as
\begin{equation}\label{eq:p_theta_t}
P(\theta,\tau) = \frac{1}{2 \pi}\int K(\theta,\theta_0,\tau) \,{\rm d}\theta_0.
\end{equation}
Our strategy is to approximate $P(\theta,\tau)$ by approximating $K$ analytically and performing the integral \eqref{eq:p_theta_t} numerically.

The propagator $K$ can be approximated in the small $\varepsilon$ limit using the semiclassical approach described in Appendix \ref{sec:app}.
In particular, we are satisfied with an approximation valid for short times, because the long-time behavior of $P(\theta,\tau)$ is suppressed in the integral \eqref{eq:rt_dist} for the steady-state distribution.
Therefore, we make use of the Gaussian approximation of $K$ about a deterministic trajectory, derived for a general 1D Fokker-Planck equation of the form \eqref{eq:fokk_theta_no_tumble} in Sec.~\ref{sec:gauss_approx}.
In this case, $K$ is peaked around the trajectory $\theta^*(\tau)$ initiated at $\theta_0$, given by Eq.~\eqref{eq:theta_t_exact}.
The final expression for $K$ follows from Eqs.~\eqref{eq:gauss_prop} and \eqref{eq:gauss_var}, which after lengthy but straightforward calculations yields
\begin{equation}\label{eq:theta_gauss}
K(\theta,\theta_0,\tau)  \approx \sqrt{\frac{1}{2\pi \varepsilon} \frac{\partial^2 R}{\partial \theta^2}}\exp\left[ - \frac{1}{2\varepsilon} \frac{\partial^2 R}{\partial \theta^2} (\theta - \theta^*(\tau))^2\right], 
\end{equation}
where
\begin{equation} \label{eq:theta_var}
\frac{\partial^2 R}{\partial \theta^2}  = \frac{4 \alpha \left(e^{2 \alpha \tau} + e^{-2 \alpha \tau} \tan^2 \theta_0 \right)^2}{e^{4\alpha \tau} + 8 \alpha \tau \tan^2\theta_0 - e^{-4 \alpha \tau} \tan^4 \theta_0 - 1 + \tan^4\theta_0}.
\end{equation}

\begin{figure}[t!]
\centering
\includegraphics[width=\textwidth]{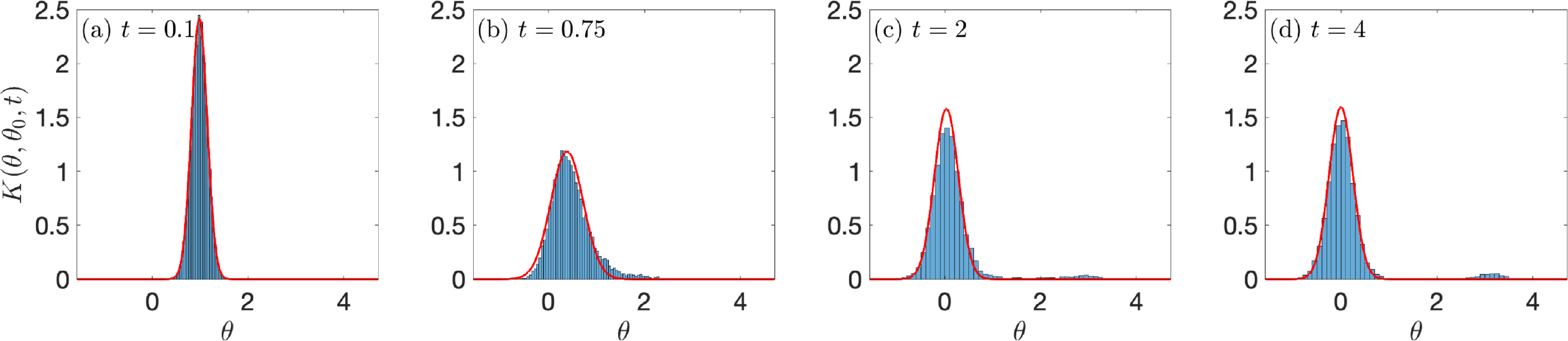}
\caption{Time evolution of the propagator $K(\theta,\theta_0,t)$ for $\theta_0 = 1.07$, $\alpha = 1$, and $\varepsilon= 0.25$. Histograms: numerical simulations of Eq.~\eqref{eq:theta_noise} with $10^4$ trajectories. Red curves: theoretical prediction given by Eq.~\eqref{eq:theta_gauss}.}\label{fig:p_theta_t}
\end{figure}
We illustrate the validity of the approximate probability distribution \eqref{eq:theta_gauss} by comparing the prediction to numerical simulations of Eq.~\eqref{eq:theta_noise}.
One comparison is shown in Fig.~\ref{fig:p_theta_t}, where at $t = 0$ we initialized the swimmers with the orientation $\theta_0 = 1.07$, not terribly far from the turning point $\theta = \pi/2 \approx 1.57$, with a modest noise strength of $\varepsilon = 0.25$.
These parameters were selected to push the limits of our approximations; not only do we assume $\varepsilon$ is small, but our calculation of $K(\theta,\theta_0,\tau)$ neglects entirely the contributions of paths that cross the turning point at $\theta = \pi/2 \approx 1.57$ and relax to the equilibrium at $\theta = \pi$ instead of $\theta = 0$.
Despite these limitations, we see that the approximate  $K(\theta,\theta_0,\tau)$ given by Eq.~\eqref{eq:theta_gauss} does a good job both of tracking the center of the distribution of trajectories and accounting for their spread as a function of time.
After some time, the variance of the approximate distribution $(\partial^2R/\partial \theta^2)^{-1}$ saturates and the centroid converges onto $\theta=0$, yielding a steady-state.
This distribution is reasonably close to the numerical one at $t = 4$.
However, we know that if we continue the numerical simulations for very long times, then eventually the distribution should approach the exact steady-state given by Eq.~\eqref{eq:theta_dist} (Figs.~\ref{fig:theta_rt_dist}a and \ref{fig:theta_rt_dist}b).
In contrast to our approximate distribution, which only has one peak that eventually converges to $\theta = 0$, the true steady-state distribution is symmetrically peaked about $\theta = 0$ and $\theta = \pi$.
Over long times, this is achieved as the noise drives some swimmers' orientations over the potential barriers at $\theta = \pm \pi/2$, causing them to settle down around $\theta = \pi$ for long times.
This process is reflected by the growing peak in the density of simulated trajectories at $\theta = \pi$ in Fig.~\ref{fig:p_theta_t}c and \ref{fig:p_theta_t}d.
Our approximate distribution manifestly neglects this process, because the action associated with such trajectories is larger than for trajectories near the deterministic path, which are the only trajectories accounted for in our approximation.

\subsubsection{Approximation of the stationary state}
\begin{figure}[t!]
\centering
\includegraphics[width=0.8\textwidth]{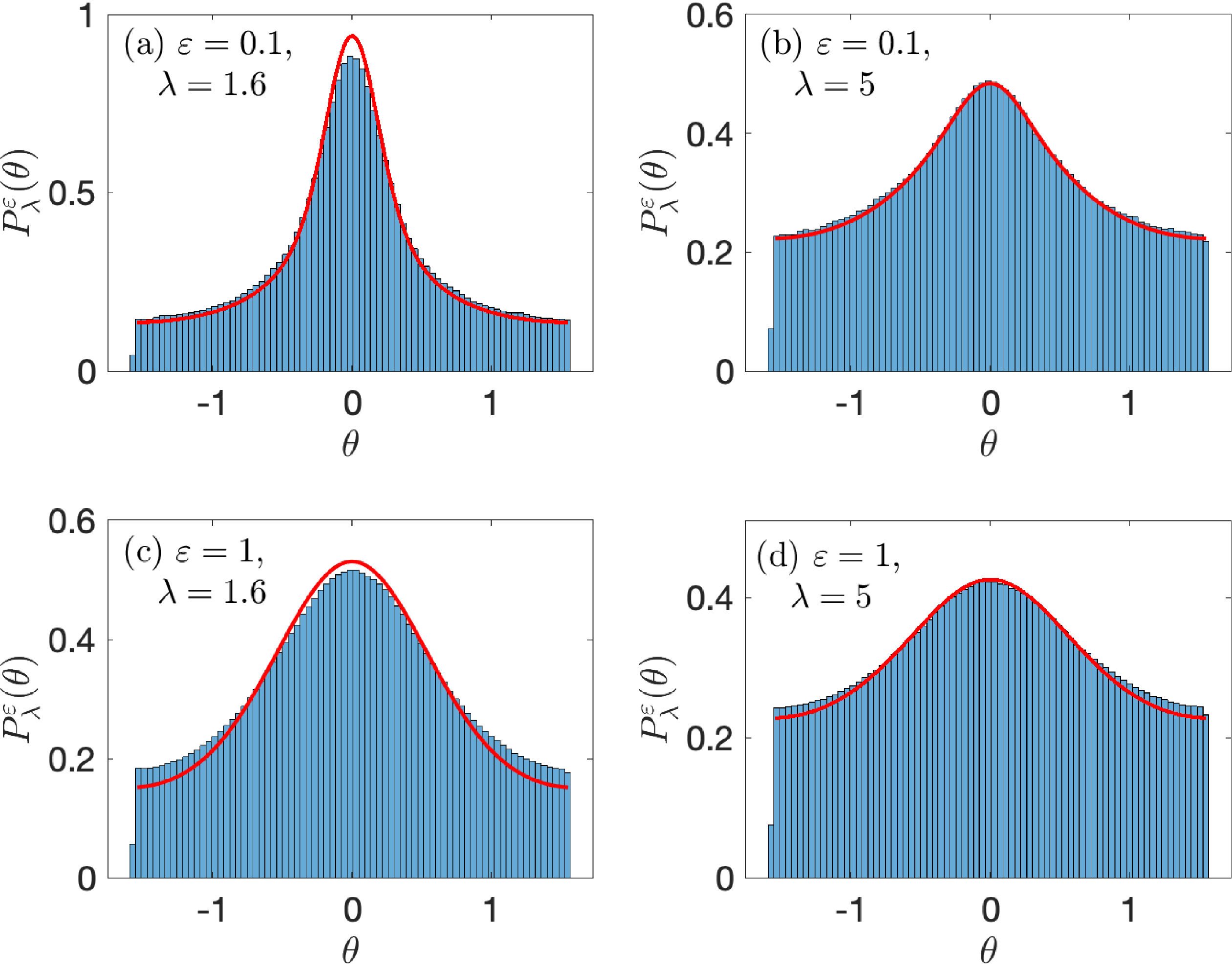}
\caption{Stationary $\theta$ distributions with $\alpha=1$, with both tumbling and rotational diffusion. Histograms are Monte Carlo simulations of Eq.~\eqref{eq:thdot}, and red curves are the theoretically predicted distributions given by the numerical evaluation of Eq.~\eqref{eq:rt_dist} using Eqs.~\eqref{eq:p_theta_t} and \eqref{eq:theta_gauss}. Distributions are plotted in the range $\theta \in (-\pi/2,\pi/2)$. (a) $\varepsilon = 0.1$, $\lambda = 1.6$. (b) $\varepsilon = 0.1$, $\lambda = 5$. (c) $\varepsilon = 1$, $\lambda = 1.6$. (d) $\varepsilon = 1$, $\lambda = 5$.}\label{fig:p_lambda_star_sig_316}
\end{figure}
We can now compute the stationary orientation distributions of swimmers with both tumbling and rotational diffusion.
To recap, we have an explicit approximation \eqref{eq:theta_gauss} of $K(\theta,\theta_0,\tau)$, the time-dependent probability distribution of $\theta$ for a  rotationally-diffusing swimmer with orientation $\theta_0$ at time $t=0$ (which also requires Eqs.~\eqref{eq:theta_var} and \eqref{eq:theta_t_exact} to evaluate).
Thus, we are able to numerically evaluate our expression \eqref{eq:p_theta_t} for the time-dependent probability distribution $P(\theta,\tau)$ of $\theta$ for a rotationally-diffusing swimmer with an initially uniform orientation distribution.
This initial state corresponds to the swimmer's orientation distribution after a tumble.
Therefore, we can finally evaluate Eq.~\eqref{eq:rt_dist} for $P_\lambda^*(\theta)$, the stationary $\theta$ distribution of a tumbling and rotationally-diffusing swimmer.

We proceed by evaluating Eqs.~\eqref{eq:p_theta_t} and \eqref{eq:rt_dist} numerically, and we compare the results with numerical simulations of tumbling and rotationally-diffusing swimmers, i.e.\ simulations of Eq.~\eqref{eq:thdot}.
The results are shown in Fig.~\ref{fig:p_lambda_star_sig_316}, with all four possible combinations of $\varepsilon$ and $\lambda$ used in Fig.~\ref{fig:theta_rt_dist}.
Without rotational diffusion (Figs.~\ref{fig:theta_rt_dist}c and \ref{fig:theta_rt_dist}d), the distribution peak at $\theta=0$ is very sharp.
Comparing with the distributions in Fig.~\ref{fig:p_lambda_star_sig_316},  we conclude rotational diffusion smooths out these peaks.
We observe good agreement between the stochastic simulations and the semiclassical theory in all cases.
Thus, our semiclassical method for evaluating $P_\lambda^\varepsilon(\theta)$ can in principle be used to fit experimental data, allowing the determination of the effective rotational diffusivity and tumbling rate of swimmers in the hyperbolic flow.

\section{Depletion effect}\label{sec:depletion}
Here, we present Monte Carlo and semiclassical calculations quantifying the depletion effect.
We quantify the depletion effect by calculating the probability $\Pr(x_0)$ that a swimmer ultimately exits right with a given initial position $- 1 < x_0 < 1$ and a given intensity of the noise.
For an $x_0$ near the BIM $x = -1$, the signature of the depletion effect is a decreasing $\Pr(x_0)$ for increasing noise intensity.
A low probability of right-exiting swimmer trajectories initialized near $x = -1$ would be consistent with the absence of such trajectories for run-and-tumble bacteria in the experimental data shown in Fig.~\ref{fig:expt}.
Conversely, for an $x_0$ near the BIM $x=1$, $\Pr(x_0)$ should increase with increasing noise intensity.
This is simply due to the symmetry of the hyperbolic flow, which requires that
\begin{equation}\label{eq:symmetry_prob}
\Pr(-x_0) = 1 - \Pr(x_0).
\end{equation}
We focus solely on the dynamics in the $x\theta$ plane, because it is independent of the $y$ variable, as discussed in Sec.~\ref{sec:hyp}.
Hence, Eq.~\eqref{eq:mastereq} becomes
\begin{equation}\label{eq:fp_xtheta}
\frac{\partial P}{\partial t} = -\nabla \cdot (\f P) + \frac{\varepsilon}{2}\left( \gamma\frac{\partial^2 P}{\partial x^2}  + \frac{\partial^2 P}{\partial \theta^2} \right)+ \lambda \left[ -P + \frac{1}{2\pi} \int_0^{2\pi} P(x,\theta',t){\rm d} \theta' \right],
\end{equation}
where
\begin{equation}
\f = (x+\cos\theta,-\alpha \sin(2\theta))
\end{equation}
is the drift restricted to the $x\theta$ plane.
In Eq.~\eqref{eq:fp_xtheta} and throughout this section, we also take $\nabla = (\partial/\partial x, \partial/\partial \theta)$.

We restrict our attention to the case where rotational diffusion dominates translational diffusion, i.e.\ $\gamma \ll 1$, and we fix $\gamma = 0.1$.
When $\gamma = 0$, all swimmers which cross the line $x = 1$ must ultimately exit right, due to the BIM at $x = 1$ blocking inward swimming particles.
Therefore, the probability to exit right may be calculated by integrating the probability current through $x=1$. 
We assume this remains approximately true for small $\gamma$.
Defining $\Pr(x_0,t)$ as the probability that a swimmer has exited right by time $t$, we have
\begin{equation}\label{eq:prRightint}
\Pr(x_0,t) = \int_{x > 1} P(x,\theta,t) {\rm d}x {\rm d} \theta.
\end{equation}
Differentiating Eq.~\eqref{eq:prRightint} with respect to time and using Eq.~\eqref{eq:fp_xtheta}, we obtain the probability current
\begin{align}
\frac{\partial \Pr}{\partial t} & = \int_{x > 1} \left\{-\nabla \cdot (\f P) + \frac{\varepsilon}{2}\left( \gamma\frac{\partial^2 P}{\partial x^2}  + \frac{\partial^2 P}{\partial \theta^2} \right)+ \lambda \left[ -P + \frac{1}{2\pi} \int_0^{2\pi} P(x,\theta',t){\rm d} \theta' \right] \right\} {\rm d} x {\rm d} \theta \nonumber \\ \label{eq:prRightDiv}
& = -\int_{x > 1} \nabla \cdot {\bf J} {\rm d} x {\rm d} \theta,
\end{align}
where 
\begin{equation}
{\bf J} = \left[\f  - \frac{\varepsilon}{2} {\mathsf D} \nabla (\ln P) \right] P
\end{equation}
is the probability current density excluding tumbling, with
\begin{equation}
{\mathsf D} = \begin{pmatrix}
\gamma & 0 \\
0 & 1
\end{pmatrix}.
\end{equation}
The tumbling contribution in Eq.~\eqref{eq:prRightDiv} vanishes upon integration over $\theta$.
Using the divergence theorem, the probability current \eqref{eq:prRightDiv} becomes
\begin{equation}
\frac{\partial \Pr}{\partial t} = \int_0^{2\pi} {\bf J}(1,\theta,t) \cdot \hat{\bf x} {\rm d} \theta,
\end{equation}
and thus the probability to exit right is given by
\begin{equation}\label{eq:fluxintegral}
\Pr(x_0) = \int_0^\infty {\rm d} t \int_0^{2\pi} {\rm d} \theta \left[1 + \cos \theta - \frac{\varepsilon \gamma }{2}\frac{\partial}{\partial x} \left(\ln P(1,\theta,t) \right) \right] P(1,\theta,t) .
\end{equation}
For $\gamma = 0$, Eq.~\eqref{eq:fluxintegral} is exact.
For small $\gamma > 0$, Eq.~\eqref{eq:fluxintegral} is an approximation, because swimmers close to the righthand side of the BIM may fluctuate over to the lefthand side due to translational diffusion.

\subsection{Monte Carlo calculations with diffusion or tumbling}\label{sec:mc}
\begin{figure}
\centering
\includegraphics[width=\textwidth]{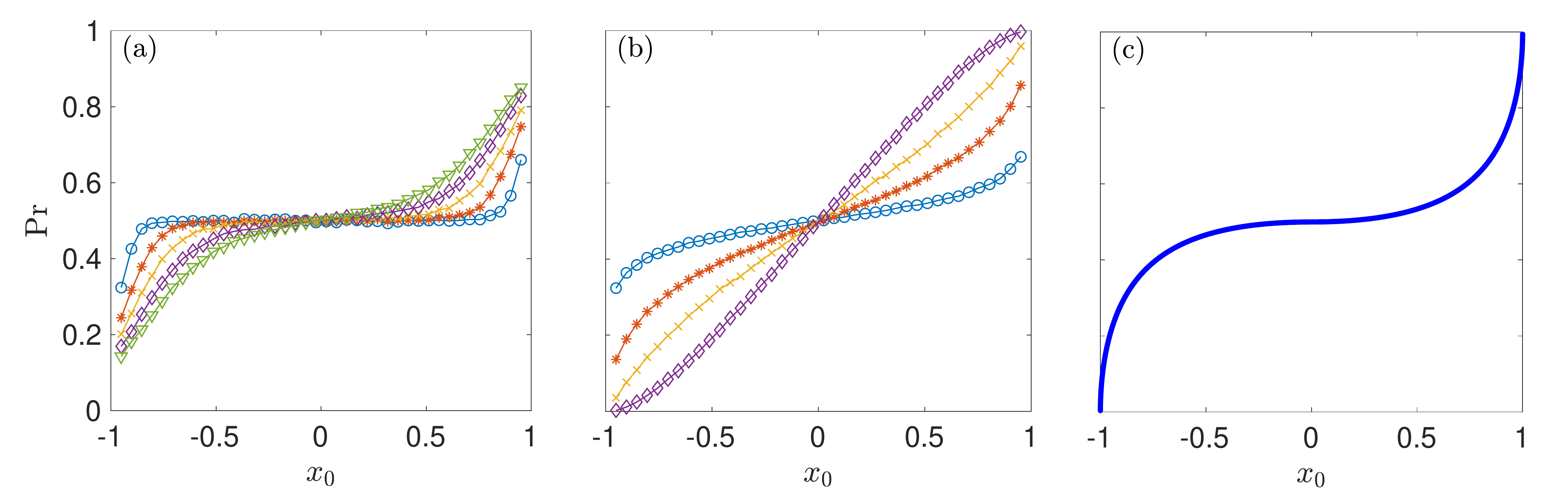}
\caption{Monte Carlo calculations of swimmer probability to exit right $\Pr(x_0)$, for $\alpha = 1$ swimmers in the hyperbolic flow. (a) $\lambda = 0$, $\gamma = 0.1$, and $\varepsilon = 0.1\,\,(\bigcirc)$, $\varepsilon = 0.3 \,\, (\ast)$, $\varepsilon = 0.5\,\, (\times)$, $\varepsilon = 0.7\,\,(\diamond)$, and $\varepsilon = 0.9\,\,(\bigtriangledown)$. (b) $\varepsilon = 0$ and $\lambda = 0.167 \,\, (\bigcirc)$, $\lambda = 0.5 \,\,(\ast)$, $\lambda = 1 \,\,(\times)$, and $\lambda = 2 \,\, (\diamond)$. (c) $\varepsilon = \lambda = 0$.}\label{fig:depletion_mc}
\end{figure}
Monte Carlo calculations of the swimmer probability to exit right as a function of $x_0$ confirm that the depletion effect is caused by noise.
For each $x_0$, we computed $\Pr(x_0)$ by integrating Eq.~\eqref{eq:hypflow} for 50,000 initial conditions with randomly selected initial orientations $\theta_0$ from $t = 0$ to $t = 6$.
The probability to exit right, according to Eq.~\eqref{eq:prRightint}, is then the fraction of trajectories for which $x > 1$ at the end of the simulation. 
Figure \ref{fig:depletion_mc} shows the results for $\Pr(x_0)$ for swimmers with diffusion only ($\lambda = 0$, Fig.~\ref{fig:depletion_mc}a) and swimmers with tumbling only ($\varepsilon = 0$, Fig.~\ref{fig:depletion_mc}b).
For the $\lambda = 0$ swimmers, $\theta_0$ was drawn from the stationary distribution $P^\varepsilon(\theta_0)$ given by Eq.~\eqref{eq:theta_dist}.
For the $\varepsilon = 0$ swimmers, $\theta_0$ was drawn from the stationary distribution $P_\lambda(\theta_0)$ given by Eq.~\eqref{eq:theta_dist_rt}.
We also show $\Pr(x_0)$ for deterministic swimmers ($\varepsilon = \lambda = 0$) initialized with a uniform distribution of $\theta_0$ in Fig.~\ref{fig:depletion_mc}c.
Here, $\Pr(x_0)$ is obtained by calculating the fraction of trajectories on the right side of the SwIM at a given $x_0$ (see Fig.~\ref{fig:swims2D}a).

Figure \ref{fig:depletion_mc} shows that as the intensity of noise increases, $\Pr(x_0)$ increases for $x_0 > 0 $ and decreases for $x_0 < 0$.
This occurs both for swimmers with diffusion only (Fig.~\ref{fig:depletion_mc}a) and for swimmers with tumbling only (Fig.~\ref{fig:depletion_mc}b), where the intensity of noise effectively increases when the tumbling frequency $\lambda$ increases.
The reduction of $\Pr(x_0)$ for $x_0 < 0$ for noisier swimmers is consistent with the depletion effect observed in the experimental data shown in Fig.~\ref{fig:expt}.
For smooth swimming bacteria (Fig.~\ref{fig:expt}a), which behave like swimmers with weak diffusion, the exit-right probability $\Pr(x_0)$ is substantial for most values of $x_0$, even those relatively close to the BIM at $x = -1$ (Fig.~\ref{fig:depletion_mc}a).
Therefore, it is not unlikely to observe bacteria trajectories which graze the BIM at $x = -1$, as we indeed see in Fig.~\ref{fig:expt}a.
For run-and-tumble bacteria on the other hand (Fig.~\ref{fig:expt}b), 
 $\Pr(x_0)$ is very small near $x_0 = -1$ for sufficiently large  $\lambda$ (Fig.~\ref{fig:depletion_mc}b).
Therefore, it is very unlikely to observe bacteria trajectories that pass near $x = -1$ and subsequently exit right, explaining the paucity of trajectories near $x=-1$ in Fig.~\ref{fig:expt}b.
Because fluctuations can cause the swimmers to cross one-way barriers in the flow, fluctuations can dramatically impact a swimmer's ability to navigate a fluid flow.

\subsection{Semiclassical approximation for diffusion}
\begin{figure}
\centering
\includegraphics[width=0.7\textwidth]{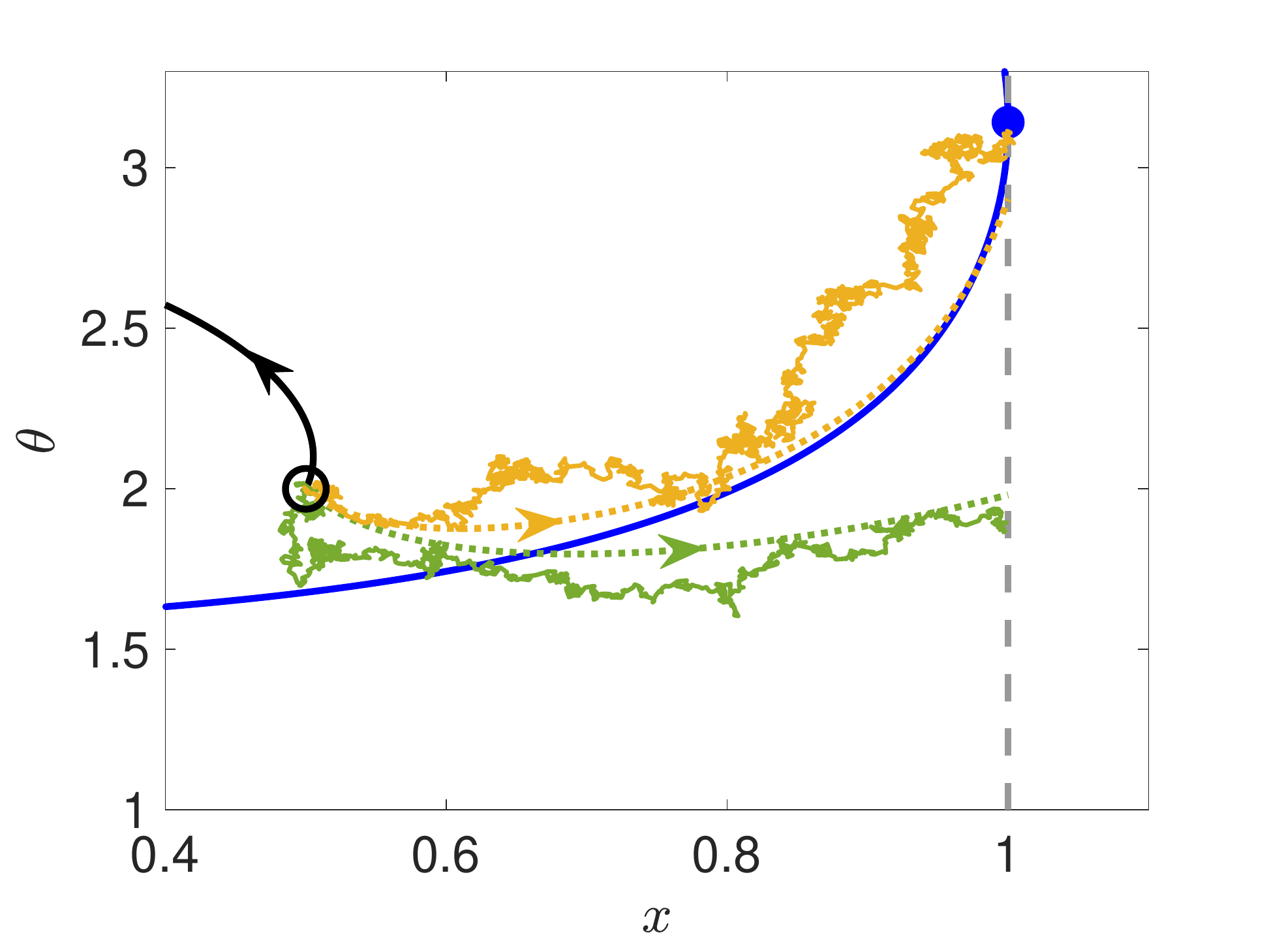}
\caption{Comparison between the minimum-action paths and noisy trajectories ($\lambda = 0$, $\gamma = 0.1$, $\varepsilon = 0.0625$) of a swimmer exiting right with an initial condition $(x_0,\theta_0) = (0.5,2)$ (black circle). The black curve is the deterministic trajectory, which exits left. The solid jagged curves  are representative noisy trajectories hitting $x = 1$ in time $t \approx 1$ (green) and $t \approx 2$ (yellow). The dotted curves are the minimum-action paths hitting $x = 1$ at $t = 1$ (green) and $t = 2$ (yellow). They are projections into the $x\theta$ plane of solutions to the boundary value problem seeking the trajectories $(x(t),\theta(t),p_x(t),p_\theta(t))$ of Hamiltonian \eqref{eq:ham_hyp} with the specified initial condition $(x_0,\theta_0)$, final position $x(t) = 1$, and final momentum $p_\theta(t) = 0$. The blue curve is the stable SwIM of the swimming fixed point (blue dot).}\label{fig:least_action}
\end{figure}
We use the semiclassical approximation to compute $\Pr(x_0)$ when $\lambda = 0$ and investigate how accurately it matches the Monte Carlo calculations.
For the $x\theta$ dynamics in the hyperbolic flow, Hamiltonian \eqref{eq:Hamiltonian} becomes
\begin{equation}\label{eq:ham_hyp}
H(x,\theta,p_x,p_\theta) = \gamma \frac{p_x^2}{2} + \frac{p_\theta^2}{2} + p_x(x + \cos \theta) - p_\theta \alpha \sin (2\theta).
\end{equation}
We evaluate Eq.~\eqref{eq:fluxintegral} for $\Pr(x_0)$ using our semiclassical approximation for $P(x,\theta,t)$.
This essentially requires integrating over a subset of trajectories of Eq.~\eqref{eq:ham_hyp}, which begin at $x_0$ at $t = 0$ and hit $x = 1$ at a later time.
One advantage of the semiclassical approximation is that this set of trajectories is independent of $\varepsilon$, so once these trajectories are computed, Eq.~\eqref{eq:fluxintegral} can be evaluated for any arbitrary value of $\varepsilon$.
Another advantage is that this set of trajectories provides insight into the actual paths in $x\theta$ space that noisy swimmers take on their way to exiting right. 

To illustrate the relationship between the trajectories of Hamiltonian \eqref{eq:ham_hyp} and the noisy trajectories, we first consider the semiclassical evolution of a probability density initially concentrated at a single point $(x_0,\theta_0)$.
This corresponds to an initial condition
\begin{equation}
P_0(x,\theta) = \delta(x - x_0)\delta(\theta - \theta_0),
\end{equation}
which is the initial condition for the propagator of the Fokker-Planck equation (Eq.~\eqref{eq:prop_ic}).
The semiclassical solution for such an initial condition (Eq.~\eqref{eq:final_prop}) requires one to integrate all trajectories of Hamiltonian \eqref{eq:ham_hyp} beginning at $(x_0,\theta_0)$, which means considering all possible initial momenta $(p_{x0},p_{\theta0})$ at that point (Appendix \ref{sec:HJ}).
This 2D surface of initial conditions of the Hamiltonian system is called a Lagrangian manifold \cite{Littlejohn1992}.
Along the way, one keeps track of the accumulated action $R(x,\theta,x_0,\theta_0,t)$ 
along each trajectory (Eqs.~\eqref{eq:Wdot} and\eqref{eq:action}).
For Hamiltonian \eqref{eq:ham_hyp}, the accumulated action is
\begin{equation}\label{eq:accumaction}
R(x,\theta,x_0,\theta_0,t) = \frac{1}{2}\int_0^t \left[\gamma p_x(\tau)^2 + p_\theta(\tau)^2\right] {\rm d} \tau,
\end{equation}
where the integral is along the trajectory connecting $(x_0,\theta_0)$ to $(x,\theta)$ in time $t$.
In the case of the propagator initial condition, the function $W$ of the semiclassical probability density \eqref{eq:WKB} is simply equal to the accumulated action, $W(x,\theta,t) = R(x,\theta,x_0,\theta_0,t)$.
The exponential dependence of the semiclassical probability density on $W$ makes the probability density peaked around the local minima and valleys of $W$.
The Hamiltonian trajectories reaching these minima or valleys can be thought of as prototypical noisy trajectories.

For example, we consider swimmers exiting right from $(x_0,\theta_0) = (0.5,2)$, as shown in Fig.~\ref{fig:least_action}.
For this initial condition, a deterministic swimmer would exit left, because it is to the left of the SwIM.
Noise allows some of the swimmers to cross over the SwIM and exit right, as illustrated by the two sample trajectories selected from a Monte Carlo simulation in Fig.~\ref{fig:least_action}.
We selected one trajectory that hits $x = 1$ at $t \approx 1$, and a second trajectory that hits at $t \approx 2$; aside from these prescribed hitting times, the trajectories were selected at random.
We can calculate the trajectories of the system with Hamiltonian \eqref{eq:ham_hyp} which hit $x = 1$ at those same times.
There are infinitely many, each hitting with a different final $\theta$.
Out of this set of trajectories, we find the ones which minimize the action at $x = 1$, equivalent to the condition
\begin{equation}\label{eq:valley}
\frac{\partial W(1,\theta,t)}{\partial \theta} = 0 = p_\theta(t),
\end{equation}
where the last equality follows from Eq.~\eqref{eq:lagman}.
In other words, for a given $t$, the minimum-action trajectory is the one which hits $x = 1$ with $p_\theta(t) = 0$.
Equation \eqref{eq:valley} is the condition for a valley of $W(x,\theta,t)$ because it is a local minimum of $W$ with $x$ (and $t$) held fixed.
The minimum-action trajectories corresponding to the two noisy trajectories are plotted as the dotted curves in Fig.~\ref{fig:least_action}.

The resemblance between the noisy paths and the minimum-action paths in Fig.~\ref{fig:least_action} demonstrates the power of the semiclassical approximation to the Fokker-Planck equation.
The deterministic trajectories underlying the semiclassical approximation predict the paths taken by the noisy system satisfying specific boundary conditions---in this case going from $(x_0,\theta_0)$ at $t = 0$ to $x = 1$ at specified times $t$.
In the asymptotic $\varepsilon \rightarrow 0$ limit, the probability density becomes increasingly concentrated along the minimum-action paths.
However, for any finite $\varepsilon$, the probability density has a finite width around these minimum-action paths, so any actual noisy trajectory will deviate from the minimum-action path, as seen in Fig.~\ref{fig:least_action}.
The minimum-action paths are thus prototypical noisy paths with given boundary conditions, in the sense that they are the peak of the distribution of noisy trajectories satisfying those boundary conditions.
Furthermore, by taking into account the full set of trajectories of Hamiltonian \eqref{eq:ham_hyp} satisfying the boundary conditions (i.e.\ not only those in the valley of the action), one can construct the full probability distribution of trajectories satisfying the boundary conditions.
This requires computing additional quantities along the Hamiltonian trajectories that are needed to evaluate the probability density prefactor $A$ in Eq.~\eqref{eq:WKB} (see Eq.~\eqref{eq:amp1} and Eq.~\eqref{eq:final_prop} for explicit expressions and Table \ref{tab}).

\begin{table}
\centering
\begin{tabular}{p{0.1\linewidth} | p{0.6\linewidth} | p{0.3\linewidth}}
Variable & Meaning & Appendix references \\ \hline
$A$ & prefactor of probability density \eqref{eq:WKB} & Eqs.~\eqref{eq:Atransport} and \eqref{eq:A_hybrid} \\ \hline
$\bq$ & $\bq = (x,\theta) = \bq(\bz',t)$, projection of Lagrangian manifold into configuration space, as a function of initial Lagrangian manifold coordinate $\bz' = (p_{x0},\theta_0)$ and time & Eq.~\eqref{eq:qQ}, \ref{sec:hybrid} \\ \hline
$\partial \bq / \partial \bz'$ & Jacobian matrix of the projection $\bq(\bz',t)$  & Eqs.~\eqref{eq:dqdp}, \eqref{eq:dqdpdot}\\ \hline
$\int_0^t \nabla \cdot \f {\rm d} \tau$ & $\nabla \cdot \f = 1  - 2 \alpha \cos(2\theta)$ for the $x\theta$ dynamics in the hyperbolic flow; integral performed along the Hamiltonian trajectory & \ref{sec:A} \\ \hline
\end{tabular}
\caption{Summary of key quantities that appear in the formulas for the semiclassical probability density.}\label{tab}
\end{table}
Next, we turn to the semiclassical calculation of $\Pr(x_0)$, given that the swimmer's initial orientation $\theta_0$ is distributed according to Eq.~\eqref{eq:theta_dist} as in Sec.~\ref{sec:mc}.
This requires the solution of the Fokker-Planck equation \eqref{eq:fp} for $P(x,\theta,t)$ with initial condition
\begin{equation}
P_0(x,\theta) = \delta(x- x_0) \left[2 \pi I_0 \left(\frac{\alpha}{\varepsilon}\right) \right]^{-1} \exp\left[\frac{\alpha \cos 2\theta}{\varepsilon}\right].
\end{equation}
This is a hybrid propagator-WKB initial condition of the form \eqref{eq:hybridic}, where, in the notation of Appendix \ref{sec:app}, $A_0 = \left[2 \pi I_0 \left(\frac{\alpha}{\varepsilon}\right) \right]^{-1}$ and $U(\theta) = - \alpha \cos 2\theta$.
We use the semiclassical probability density \eqref{eq:final_hybrid} to evaluate Eq.~\eqref{eq:fluxintegral}.
This means that for each $x_0$, we must integrate over all Hamiltonian paths which hit $x = 1$, with initial conditions on the Lagrangian manifold
\begin{equation}\label{eq:lagmanPr}
\left\{ (x_0,\theta_0,p_{x0},p_{\theta0})\,\, \bigg| \,\, \forall\,\, \theta_0,\, p_{x0},\, p_{\theta0} \,\,\,\text{such that}\,\,\, p_{\theta0} = \frac{\partial U}{\partial \theta}(\theta_0) \right\}.
\end{equation}
Therefore, the initial Lagrangian manifold may be parametrized by the coordinates $\bz' = (p_{x0},\theta_0)$.
The probability current integral Eq.~\eqref{eq:fluxintegral} becomes
\begin{align}
\Pr(x_0) & = \frac{1}{\sqrt{2\pi\varepsilon}} \left[2 \pi I_0 \left(\frac{\alpha}{\varepsilon}\right) \right]^{-1} \int_0^\infty {\rm d} t \int_{x = 1} {\rm d} \theta \left[1 + \cos \theta + \frac{\gamma}{2}p_x - \frac{1}{2}\varepsilon \gamma \frac{\partial}{\partial x} (\ln A(1,\theta,t)) \right] \times \nonumber\\ \label{eq:fluxintegral_semi}
 &  \left|\det \frac{\partial \bq}{\partial \bz'}\right|^{-1/2} \exp \left[-\frac{(U(\theta_0) + R(1,\theta,x_0,\theta_0,t))}{\varepsilon}-\frac{1}{2} \int_0^t \nabla \cdot \f {\rm d} \tau \right],
\end{align}
where $A$ is given by Eq.~\eqref{eq:A_hybrid} and $R$ is given by Eq.~\eqref{eq:accumaction}.
Equation \eqref{eq:fluxintegral_semi} must be integrated over the set of $\theta$ and $t$ values at which the trajectories of Hamiltonian \eqref{eq:ham_hyp} hit $x = 1$.
The meaning of the new variables introduced in Eq.~\eqref{eq:fluxintegral_semi}, along with references to the appendix, is summarized in Table \ref{tab}. 

\begin{figure}
\centering
\includegraphics[width=0.6\textwidth]{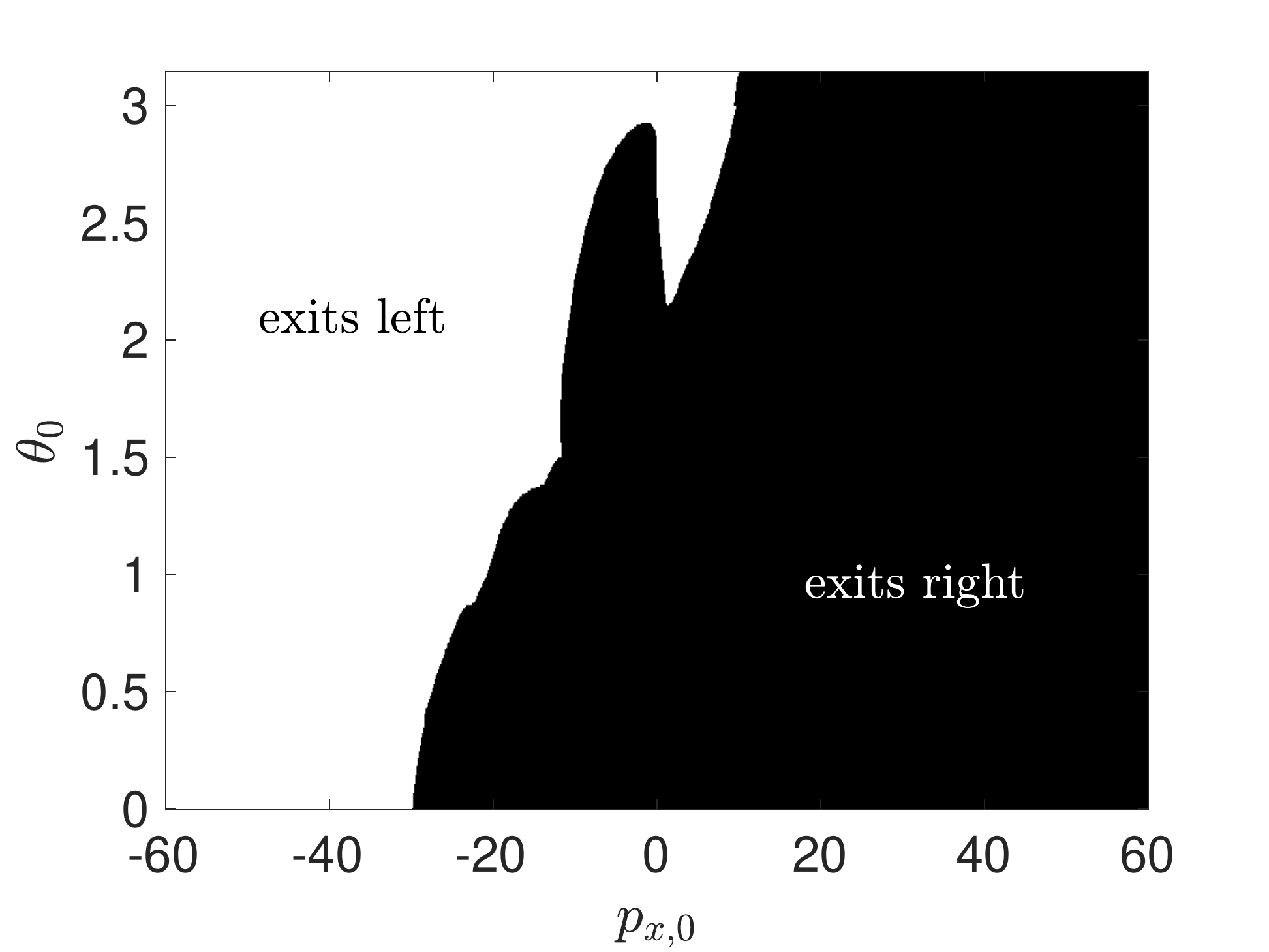}
\caption{Integration domain for Eq.~\eqref{eq:fluxintegral_approx} for $x_0$ = 0.5. The quantity $p_{x0}$ is the initial condition of the canonically conjugate momentum to $x$, and $\theta_0$ is the initial orientation of the swimmer. The initial conditions in the black region eventually hit $x = 1$, and hence Eq.~\eqref{eq:fluxintegral_approx} is integrated over the black region only. Initial conditions in the white region hit $x = -1$ instead.}\label{fig:intdomain}
\end{figure}
We make some modifications to Eq.~\eqref{eq:fluxintegral_semi} before evaluating it numerically.
The integral over final coordinates $(\theta,t)$ can be converted to an integral over the initial coordinates of the Lagrangian manifold $(p_{x0},\theta_0)$ using the Jacobian determinant
\begin{equation}
{\rm d} \theta {\rm d} t = \left| \frac{\det(\frac{\partial \bq}{\partial \bz'})}{F_x + \gamma p_x} \right| {\rm d}p_{x0}{\rm d} \theta_0 = \left| \frac{\det(\frac{\partial \bq}{\partial \bz'})}{x + \cos \theta + \gamma p_x} \right| {\rm d}p_{x0}{\rm d} \theta_0.
\end{equation}
This converts Eq.~\eqref{eq:fluxintegral_semi} into an initial value representation \cite{Heller1991,Miller1991}.
We also neglect the $\partial (\ln A)/\partial x$ term at the end of the first line of Eq.~\eqref{eq:fluxintegral_semi}, because it is of order $\varepsilon$ relative to the other terms.
This means it is of higher order in $\varepsilon$ than we account for in our asymptotic expression Eq.~\eqref{eq:WKB} (see also Eq.~\eqref{eq:pertexp}), and thus it may be neglected within the framework of the semiclassical approximation.
We must also truncate the range of $p_{x0}$ for numerical evaluation, so we take $|p_{x0}| < p_{\max}$.
Lastly, Eq.~\eqref{eq:fluxintegral_semi} is even in $\theta_0$ by symmetry, so we can restrict the domain $\theta_0 \in [0,\pi]$ and double the result.
Hence, the final expression that we evaluate numerically is 
\begin{align}
\Pr(x_0) & \approx \frac{2}{\sqrt{2\pi\varepsilon}} \left[2 \pi I_0 \left(\frac{\alpha}{\varepsilon}\right) \right]^{-1} \int_{-p_{\max}}^{p_{\max}} {\rm d} p_{x0} \int_0^\pi  {\rm d} \theta_0 \,  \frac{1 + \cos \theta + \frac{\gamma}{2}p_x}{|1 + \cos \theta + \gamma p_x|} \times \nonumber\\ \label{eq:fluxintegral_approx}
 &  \left|\det \frac{\partial \bq}{\partial \bz'}\right|^{1/2} \exp \left[-\frac{(U(\theta_0) + R(1,\theta,x_0,\theta_0,t))}{\varepsilon}-\frac{1}{2} \int_0^t \nabla \cdot \f {\rm d} \tau \right].
\end{align}
The domain for the integral \eqref{eq:fluxintegral_approx} is the set of initial conditions $(p_{x0},\theta_0)$ that eventually exit right, i.e.\ those initial conditions that reach $x = 1$ at some time $t$.
Figure \ref{fig:intdomain} shows an example integration domain for $x_0 = 0.5$.

\begin{figure}
\centering
\includegraphics[width=0.7\textwidth]{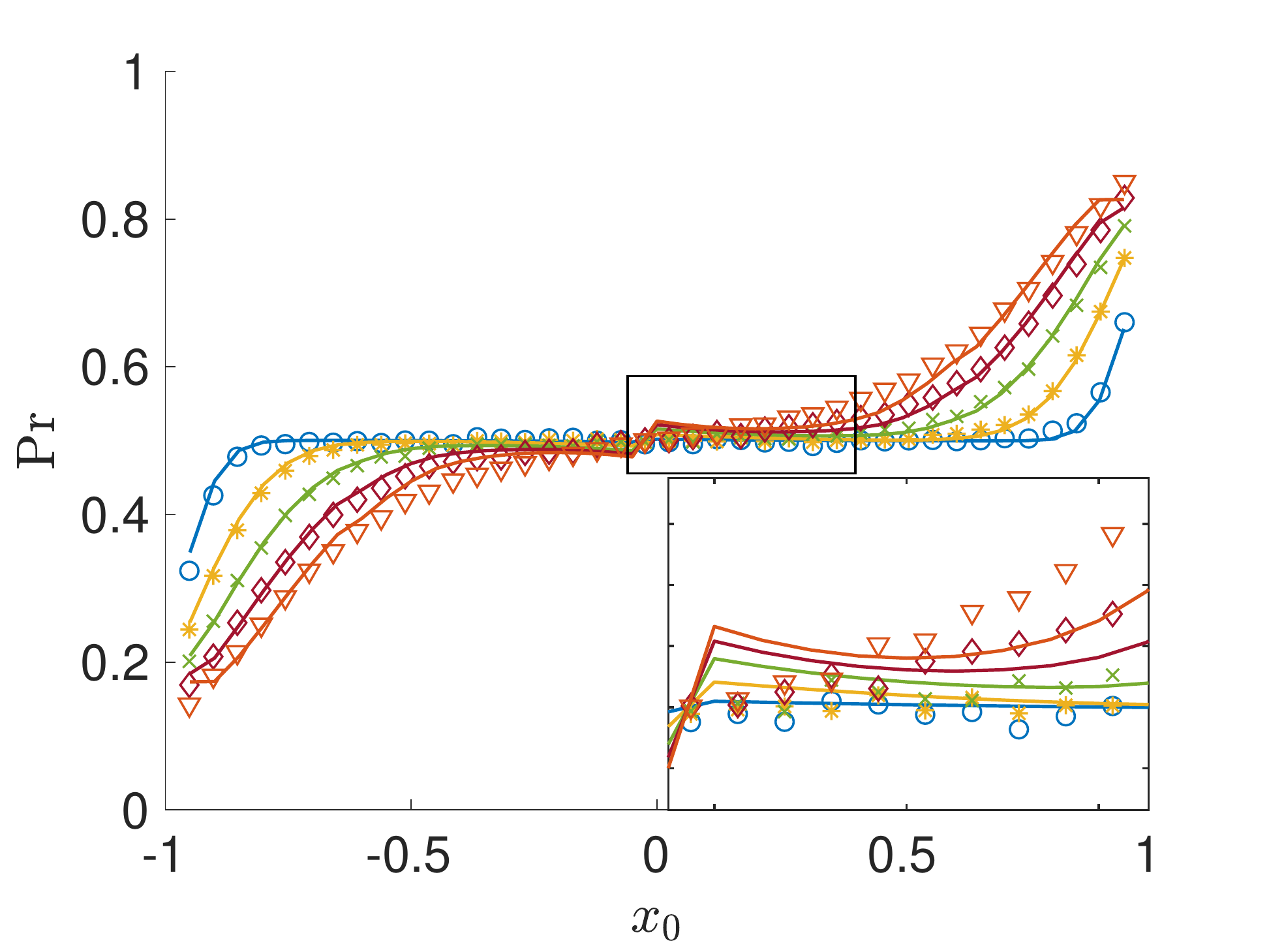}
\caption{Comparison of Monte Carlo and semiclassical calculations of $\Pr(x_0)$, for $\alpha = 1$, $\gamma = 0.1$, and $\lambda = 0$. The solid lines are the semiclassical predictions, while markers are the Monte Carlo calculations. $\varepsilon = 0.1\,\,(\bigcirc)$, $\varepsilon = 0.3 \,\, (\ast)$, $\varepsilon = 0.5\,\, (\times)$, $\varepsilon = 0.7\,\,(\diamond)$, and $\varepsilon = 0.9\,\,(\bigtriangledown)$.}\label{fig:mc_semi}
\end{figure}
We evaluate Eq.~\eqref{eq:fluxintegral_approx} numerically using the trapezoidal rule.
For each $x_0$, we discretize the set of initial conditions $(p_{x0},\theta_0)$ on the Lagrangian manifold \eqref{eq:lagmanPr} with a uniform grid.
We numerically integrate each trajectory until it hits $x=1$, up to a maximum integration time of $t = 6$, consistent with the corresponding Monte Carlo calculations in Fig.~\ref{fig:depletion_mc}a.
We simultaneously calculate the accumulated action $R$ (Eq.~\eqref{eq:accumaction}), the integral $\int \nabla \cdot \f {\rm d} \tau$, and the Jacobian matrix $\partial \bq/\partial \bz'$.
This last step requires integrating the tangent flow along the trajectories (Eq.~\eqref{eq:dqdpdot}).
The integral \eqref{eq:fluxintegral_approx} is then evaluated for a given $\varepsilon$ by summing over all the trajectories that hit $x=1$, with all of the quantities in the integrand evaluated at that moment.
Note that once the set of trajectories and all auxiliary quantities are obtained for a given $x_0$, Eq.~\eqref{eq:fluxintegral_approx} can be evaluated for an arbitrary $\varepsilon$.
This represents one of the chief advantages of the semiclassical approximation.
We take $p_{\max} = 60$.
Including higher values of $p_{x0}$ has a negligible effect on the results, because trajectories with larger $p_x$ have a larger accumulated action $R$ (Eq.~\eqref{eq:accumaction}) and thus are exponentially suppressed in Eq.~\eqref{eq:fluxintegral_approx}.
Using this approach, we calculate $\Pr(x_0)$ for a discrete set of values $x_0 \in [0,1)$, and we use the symmetry \eqref{eq:symmetry_prob} to get $\Pr(x_0)$ for $x_0 \in (-1,0)$.
The results are plotted in Fig.~\ref{fig:mc_semi}.

\subsection{Discussion}
For $|x_0|$ near $1$, we see  in Fig.~\ref{fig:mc_semi} an excellent agreement between the semiclassical predictions for $\Pr(x_0)$ and the Monte Carlo simulations.
As $\varepsilon$ increases from $0.1$ to $0.9$, we see the semiclassical predictions overlap with the Monte Carlo simulations for $x_0$ near the BIMs.
This is particularly impressive, because in addition to the small-$\varepsilon$ assumption manifest in the semiclassical model, we have made a couple additional approximations in evaluating Eq.~\eqref{eq:fluxintegral_semi}.
Therefore, by summing over all Hamiltonian trajectories that exit right, weighted appropriately by the semiclassical probability density in Eq.~\eqref{eq:fluxintegral_approx}, we can accurately calculate exit-right probability $\Pr(x_0)$.
The integral \eqref{eq:fluxintegral_approx} includes all trajectories with a given $x_0$ that exit right, including those which begin to the right side of the SwIM (Fig.~\ref{fig:swims2D}a) and would have exited right even without noise, as well as those which begin on the left side of the SwIM and cross it due to fluctuations (Fig.~\ref{fig:least_action}).

As $|x_0|$ gets closer to $0$, however, the semiclassical predictions begin to deviate from the Monte Carlo calculations with increasing $\varepsilon$.
In particular, Eq.~\eqref{eq:symmetry_prob} requires that $\Pr(0) = 0.5$, that is, a swimmer starting in the middle of the flow has an equal probability of going left or right.
While the semiclassical prediction appears consistent with this property for $\varepsilon = 0.1$, as $\varepsilon$ increases further, we see the semiclassical $\Pr(0)$ increase above $0.5$ in the inset of Fig.~\ref{fig:mc_semi}.
This fact, combined with our use of Eq.~\eqref{eq:symmetry_prob} to obtain the semiclassical $\Pr(x_0)$ for $x_0 < 0$, causes the apparent kink at $x_0 = 0$ in our semiclassical predictions plotted in Fig.~\ref{fig:mc_semi}.

\begin{figure}
\centering
\includegraphics[width=0.7\textwidth]{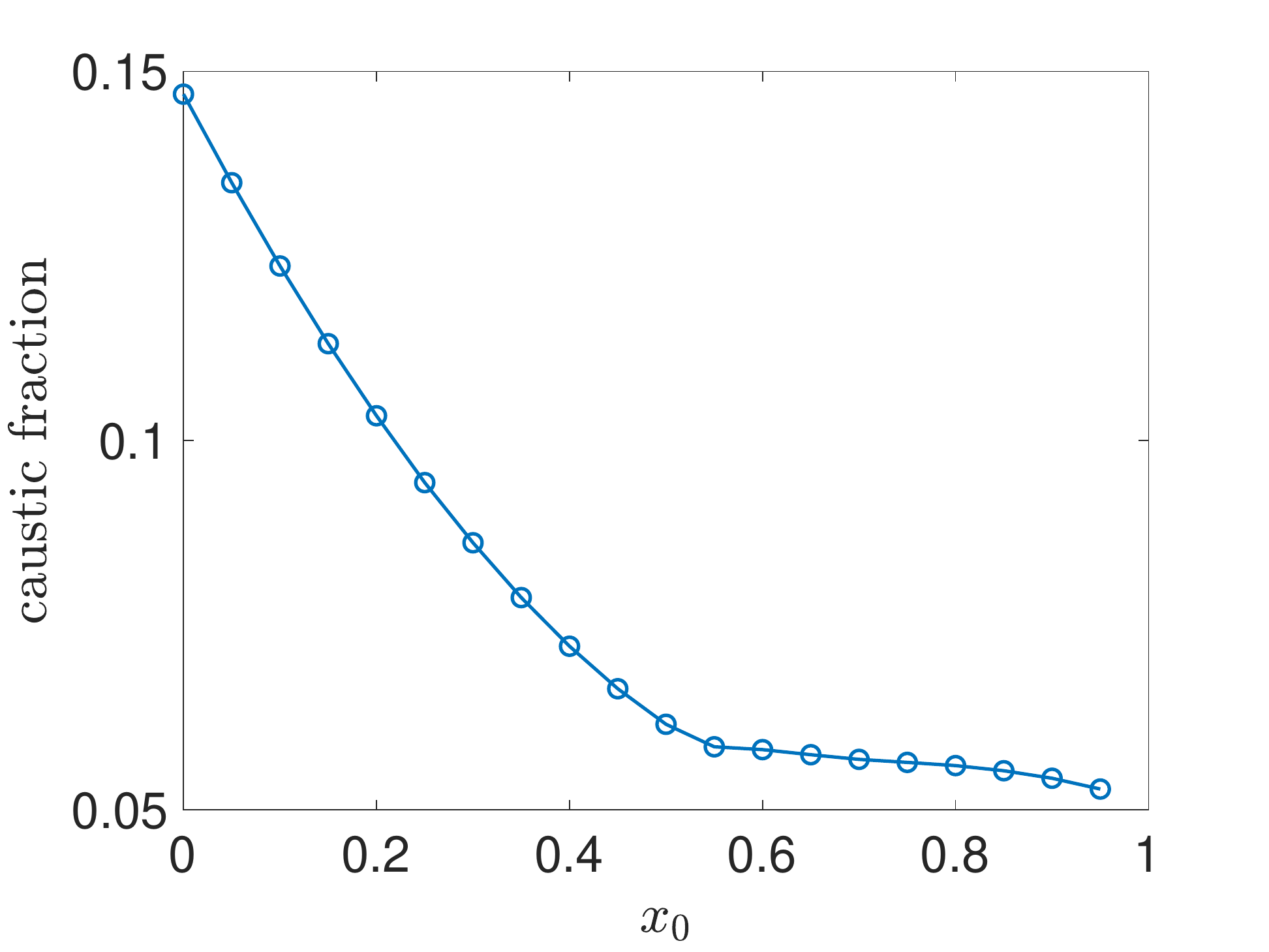}
\caption{Fraction of trajectories used in the semiclassical calculation that pass through at least one caustic before hitting $x = 1$.}\label{fig:caustic_fraction}
\end{figure}
We believe that at least part of the discrepancy between the semiclassical $\Pr(x_0)$ and the Monte Carlo calculations for $x_0$ close to 0 is due to the presence of caustics, a technical issue that we have ignored until now.
Our semiclassical approximation assumes the uniqueness of Hamiltonian trajectories that originate on the initial Lagrangian manifold and go from $(x_0,\theta_0)$ to $(x,\theta)$ in time $t$.
This means, for example, that when evaluating the accumulated action $R(x,\theta,x_0,\theta_0,t)$ in Eq.~\eqref{eq:accumaction}, there is a unique such trajectory.
The uniqueness is the critical property that makes $R$ a well-defined function.
However, uniqueness is guaranteed only for sufficiently short times, meaning that once $t$ is sufficiently large, there will be multiple Hamiltonian trajectories connecting the two points.
In this case, $R$ becomes multi-valued \cite{ArnoldCM}.
Geometrically, the uniqueness breaks down when the evolving Lagrangian manifold develops fold singularities, such that when it is projected into $\bq$ space, parts of the projection overlap with each other.
These overlap regions are the regions where multiple Hamiltonian trajectories can arrive at a single point.
In the context of the semiclassical approximation in quantum mechanics, these fold singularities are called caustics, and they have been extensively studied (see \cite{Littlejohn1992} and references therein).
The formulas for the semiclassical wave function need to be corrected to account for the occurrence of caustics through the inclusion of a Maslov index, a phase factor that essentially counts the number of caustics encountered by each trajectory.

We know of no general prescription for dealing with caustics in the semiclassical approximation to the Fokker-Planck equation, even though they commonly occur.
Previously, caustics have been investigated in the semiclassical formulation of the noise-driven dynamics of a nonlinear oscillator in two dimensions \cite{Dykman1994a}.
When calculating the steady-state probability density of this system, one must account for switching lines in phase space, i.e.\ curves on either side of which the least-action path changes discontinuously due to the presence of multiple Hamiltonian trajectories arriving at those locations.
Near the switching line, where the distinct paths with coincident endpoints have nearly the same action, the semiclassical probability density needs to account for each of the distinct paths; in fact, the signature of multiple paths leading to the same point in phase space has been observed  experimentally in a noisy electronic oscillator  \cite{Dykman1996}.
Caustics also arise in the theoretical description of noise-induced transitions in non-gradient dynamical systems with metastable fixed points \cite{Maier1992,Maier1996}.
Specifically, the quasi-stationary probability densities underlying escape from the  metastable fixed points may be approximated semiclassically, though one must go beyond the standard WKB approximation in the vicinity of the caustics.
It is not obvious how to generalize these previously obtained results to the noisy dynamics of the swimmer in the hyperbolic flow, because the swimmer phase space does not possess stable fixed points and is fundamentally transient, i.e.\ time-dependent.

Such a generalization is needed because caustics do indeed occur in the dynamics of noisy swimmers in the hyperbolic flow.
To demonstrate this, we track the number of caustics encountered by the trajectories underlying our semiclassical calculation of $\Pr(x_0)$ up until the point that they hit $x=1$.
At points where caustics occur, the projection from the Largrangian manifold into $\bq$ space is not invertible, meaning the Jacobian determinant $\det \partial \bq/\partial \bz'$ must be zero at that point.
Thus, we count the number of caustics encountered along a trajectory by tracking zero-crossings of $\det \partial \bq/\partial \bz'$.
In Fig.~\ref{fig:caustic_fraction}, we plot the fraction of trajectories used in our numerical semiclassical approximation that encounter at least one caustic on the way to $x = 1$.
While the fraction of caustic-crossing trajectories is around $6\%$ or smaller for $x_0 > 0.5$, it rapidly rises to nearly $15\%$ as $x_0$ decreases from $0.5$ to $0$.
This trend is reasonable, because caustics only begin to occur after a sufficiently long time.
Trajectories beginning closer to $x=1$ will tend to reach it sooner, potentially before many caustics have occurred.
At the same time, in the $x_0 > 0.5$ range, we see  good agreement between the semiclassical and Monte Carlo calculations in Fig.~\ref{fig:mc_semi}, while in the $x_0 < 0.5$ range we observe a discrepancy with increasing $\varepsilon$.
This correlation is evidence that the semiclassical approximation works well when few trajectories have encountered caustics, while a discrepancy can be caused by improper treatment of the caustics, which is important when considering sufficiently long-time processes.

\section{Conclusion}\label{sec:conc}
To summarize, we have quantified the effect of noise on swimmer dynamics in a steady, two-dimensional hyperbolic fluid flow.
In such a flow, swimmers are ultimately forced to escape to the left or the right, with their transient dynamics near the passive unstable fixed point determining which way they go.
Without noise, a swimmer's fate is sealed based on its position relative to the SwIM in the $x\theta$ phase space.
With noise, the swimmer's motion is a stochastic process.
We calculated the steady-state orientation distributions of diffusive, run-and-tumble, or mixed swimmers in the hyperbolic flow.
The fluctuations give some swimmers greater opportunity to cross the SwIM and exit on the opposite side than they would have without noise. 
There is however a maximal distance that swimmers can get on either side of the passive fixed point and still be able to swim back to the other side---this is where the stable BIMs block inward swimming particles.

Fluctuations make it increasingly likely that a swimmer close to one of these BIMs does indeed end up crossing it, causing irreversible changes to the fluctuating swimmers' trajectories (assuming negligible translational diffusion).
We quantified this probability using Monte Carlo calculations and a semiclassical approximation to the swimmer Fokker-Planck equation.
The semiclassical approximation accurately predicts the probability $\Pr(x_0)$ a swimmer exits right given that it began at a position $x_0$ relative to the passive fixed point, especially for $x_0$ close to the BIM.
It also predicts the probability distribution of paths that fluctuating swimmers take in phase space given specified boundary conditions.
SwIMs and BIMs are present in nonlinear flows as well, such as alternating vortex flows \cite{Berman2021a}.
Thus, we expect the depletion effect to occur in the vicinity of the BIMs of such flows as well.

This study demonstrates the utility of the semiclassical approximation for understanding the noisy dynamics of a non-trivial active matter system.
However, it also reveals a key shortcoming of the existing semiclassical theory for Fokker-Planck dynamics.
In particular, a general approach is needed for taking into account the occurence of caustics, i.e.\ multiple branches of Hamiltonian paths connecting points in configuration space.
While this issue has been examined in a few specific cases \cite{Dykman1994a,Dykman1996,Maier1992,Maier1996}, no general theory is currently available to the best of our knowledge.
A procedure for coherently summing the contributions of multiple paths, similar to the Maslov theory in quantum mechanics \cite{Littlejohn1992,Maslov1981}, would be highly desirable, both for accurate numerical computations and for the theoretical analysis of most-likely noisy paths of a dynamical system.

Finally, the semiclassical approximation may be a valuable tool for analyzing experimental data of  noisy swimmers in fluid flows.
For example, with a sufficiently large number of experimentally-recorded trajectories of the type shown in Fig.~\ref{fig:expt}, it would be possible to test the semiclassical predictions of the exit-right probability $\Pr(x_0)$.
It should also be possible to investigate the distribution of experimentally-measured trajectories satisfying specific boundary conditions \cite{Dykman1994a,Gladrow2021}.
The semiclassically-predicted distributions may be used to fit the experimental data in order to extract physical parameters, such as rotational diffusivity and swimmer shape \cite{Junot2021}.

\section*{Acknowledgments}
We thank Tom Solomon for stimulating discussions.
This material is based upon work supported by the National Science Foundation under Grant No. CMMI-1825379.

\appendix
\section{Semiclassical approximation for the Fokker-Planck equation}\label{sec:app}
We consider the stochastic process in a $d$-dimensional phase space
\begin{align}\label{eq:stochastic}
{\rm d} \bq = \left[ \bF(\bq) + \varepsilon {\bf G}(\bq) \right] {\rm d} t + \sqrt{\varepsilon}  \mathsf{C} {\rm d} {\bf w}
\end{align}
where ${\bf w} = (w_1,w_2,\ldots,w_n)$ is a set of uncorrelated Wiener processes and $\mathsf{C}$ is a $d \times n$ matrix, assumed to be constant for simplicity.
The ${\bf G}$ term is included in Eq.~\eqref{eq:stochastic} to account for potential noise-induced drift \cite{Gaspard2002}.
The corresponding Fokker-Planck equation for the probability density $P(\bq,t)$ is 
\begin{equation}\label{eq:fp}
\frac{\partial P}{\partial t} = -\nabla \cdot \left[(\bF + \varepsilon \bG) P \right] + \frac{1}{2} \varepsilon \mathsf{D} : \frac{\partial^2 P}{\partial \bq^2},
\end{equation}
where $\mathsf{D} = \mathsf{C} \mathsf{C}^{\rm T}$ is the diffusion tensor (up to a factor of $2$).
The diffusion tensor is required to be positive-definite.

Our goal is to find an approximate solution to Eq.~\eqref{eq:fp}  in the weak-noise ($\varepsilon \ll 1$) limit.
We use an approach closely related to the semiclassical approximation of quantum mechanics \cite{Littlejohn1992},  and our derivation closely follows Ref.~\cite{Gaspard2002}.
Similar techniques have been applied to stochastic dynamics in a variety of settings \cite{Graham1984,Freidlin2012,Nolting2016,Bonnemain2019,Dykman1994a,Dykman1996,Maier1992,Maier1996}.
We consider an asymptotic expansion of the solution
\begin{equation}\label{eq:pertexp}
P(\bq,t) \approx \exp \left[ -\sum_{n = 0}^{N - 1} S_n(\bq,t) \varepsilon^{n -1} \right],
\end{equation}
where $N$ is the maximum number of terms in the expansion and the $S_n$ are functions to be determined.
We restrict our attention to $N=2$ and rewrite the solution as
\begin{equation}\label{eq:propsemi1}
P(\bq,t) \approx A(\bq,t) e^{-W(\bq,t)/\varepsilon},
\end{equation}
where $W = S_0$ and $A = e^{-S_1}$.
Equation \eqref{eq:propsemi1} constitutes a WKB approximation to Eq.~\eqref{eq:fp}.
It is in the same spirit as the semiclassical approximation to the Schr\"odinger equation; with the substitution $\varepsilon \rightarrow i\hbar$, the semiclassical wave function is expressed as $\psi = Ae^{iW/\hbar}$.
In that case, $W$ is the action of the classical system associated to the quantum Hamiltonian, and $A^2 = |\psi|^2$ is the probability density of the system.
In the Fokker-Planck case, we shall see that $W$ also corresponds to the action of a particular classical Hamiltonian derived from the Fokker-Planck equation, while $A$ is essentially a normalization function.

The functions $A$ and $W$ (equivalently the $S_n$) are determined by substituting Eq.~\eqref{eq:pertexp} into Eq.~\eqref{eq:fp}.
This yields the equation
\begin{align}\label{eq:consist}
 -\sum_{n=0}^{N-1} \frac{\partial S_n}{\partial t} \varepsilon^n  = & -\left(\nabla \cdot \bF \right) \varepsilon + \bF \cdot \sum_{n=0}^{N-1} \nabla S_n \varepsilon^n  - \left(\nabla \cdot \bG\right) \varepsilon^2 + \bG \cdot \sum_{n=0}^{N-1} \nabla S_n \varepsilon^{n+1}  \nonumber \\ 
& +\frac{1}{2} \left[ - \sum_{n=0}^{N-1} \mathsf{D} : \frac{\partial^2 S_n}{\partial \bq^2} \varepsilon^{n+1} + \sum_{n,m=0}^{N-1} \nabla S_m \cdot \mathsf{D} \nabla S_n \varepsilon^{n+m} \right].
\end{align}
The goal is to equate terms of equal order in $\varepsilon$ in Eq.~\eqref{eq:consist}, which leads to equations for the $S_n$.
We are only interested in the $\varepsilon^0$ and the $\varepsilon^1$ terms.

The solutions to Eq.~\eqref{eq:consist} depend on the initial condition $P_0(\bq) \equiv P(\bq,0)$.
We derive the solutions for three types of initial conditions.
The first is a Dirac $\delta$ function centered on an arbitrary phase-space point $\bq_0$, where we use the $``0"$ subscript to refer to an initial condition fixed at some particular value.
In this case, the solution is denoted $P = K(\bq,\bq_0,t)$, where $K$ is called the propagator, with initial condition
\begin{equation}\label{eq:prop_ic}
K(\bq,\bq_0,0) = \delta(\bq - \bq_0). \\
\end{equation}
The propagator encodes the probability for the system to make a transition from $\bq_0$ to $\bq$ in time $t$.
This fact, combined with the linearity of the Fokker-Planck equation, allows its solution with any initial probability distribution $P_0$ to be written as
\begin{equation}\label{eq:prop}
P(\bq,t) = \int K(\bq,\bq_0,t) P_0(\bq_0) {\rm d}^d \bq_0.
\end{equation}
However, evaluating Eq.~\eqref{eq:prop} in practice is very computationally costly, which leads us to consider initial conditions which vary smoothly over some or all of the $\bq$ coordinates.
In particular, we consider an initial condition already in WKB form \eqref{eq:propsemi1} \cite{Bonnemain2019}, i.e.\
\begin{equation}\label{eq:WKBic}
P_0(\bq) = A_0(\bq)e^{-W_0(\bq)/\varepsilon}.
\end{equation}
Last,  we consider initial conditions that are a hybrid between the WKB form \eqref{eq:WKBic} and the propagator form \eqref{eq:prop_ic}, of the type
\begin{equation}\label{eq:hybridic}
P_0(\bq,\bq_{f0}) = \delta(\bq_f - \bq_{f0}) A_0(\bq_k)e^{-U(\bq_k)/\varepsilon}.
\end{equation}
Here, the variables are split into two groups, $\bq = (\bq_f,\bq_k)$, where the $\bq_f$ variables are fixed at a specific value $\bq_{f0}$ at $t=0$ by the $\delta$ function in Eq.~\eqref{eq:hybridic}, while the $\bq_k$ variables are distributed according to the WKB part of Eq.~\eqref{eq:hybridic}.
We are able to solve for $W$ and $A$ for this hybrid initial condition in the special case that $\mathsf{D}$ is block-diagonal, with one block corresponding to the $\bq_f$ variables and the other to the $\bq_k$ variables, i.e.\
\begin{equation}
\mathsf{D} = \begin{pmatrix}
\mathsf{D}_f & {\bf 0}_{d_f \times d_k} \\
{\bf 0}_{d_k \times d_f} & \mathsf{D}_k
\end{pmatrix},
\end{equation}
where $d_f$ is the size of $\bq_f$ and $d_k$ is the size of $\bq_k$.
The equations satisfied by $W$ and $A$ in Eq.~\eqref{eq:propsemi1} are the same in each of the three cases, but the solutions must be selected such that the initial condition is satisfied.
In Secs.~\ref{sec:HJ}--\ref{sec:norm}, we derive the solutions for the propagator initial condition \eqref{eq:prop_ic}, and in Sec.~\ref{sec:wkb}, we derive the solutions for the WKB and hybrid initial conditions.

\subsection{Hamilton-Jacobi equation}\label{sec:HJ}
From the zeroth order of Eq.~\eqref{eq:consist}, we find $W$ satisfies
\begin{align}\label{eq:HJ}
\frac{\partial W}{\partial t} & = -H\left(\bq,\nabla W\right), \\ \label{eq:ham}
H(\bq,\bp) & = \frac{1}{2}\bp \cdot \mathsf{D} \bp + \bp \cdot \bF(\bq).
\end{align} 
Equation \eqref{eq:HJ} is the Hamilton-Jacobi equation for a Hamiltonian system in a phase space of doubled dimension $2d$, with coordinates $(\bq,\bp)$ and Hamiltonian $H$ given by Eq.~\eqref{eq:ham}.
Here, $\bq$ is the coordinate of the original stochastic dynamical system \eqref{eq:stochastic}, and $\bp$ is the momentum canonically conjugate to $\bq$.
The solution of Eq.~\eqref{eq:HJ} is obtained by using the method of characteristics, which aims to find solutions of the form $W(\bq(t),t)$ along particular paths (the characteristics) $\bq(t)$.
For the case of Eq.~\eqref{eq:HJ}, the characteristics turn out to be the projections of  trajectories $(\bq(t),\bp(t))$ of the Hamiltonian system into configuration space.
The relationship between the canonical momentum and $W$ is
\begin{equation}\label{eq:lagman}
\bp = \nabla W(\bq,t).
\end{equation}
The characteristics obey the equations
\begin{align}\label{eq:qdot}
\dot{\bq} & = \frac{\partial H}{\partial \bp} = \bF(\bq) + \mathsf{D}\bp, \\ \label{eq:pdot}
\dot{\bp} & = -\frac{\partial H}{\partial \bq}  = - \frac{\partial \bF}{\partial \bq} \bp, \\ \label{eq:Wdot}
\dot{W} & = \bp \cdot \dot{\bq} - H(\bq,\bp) = \frac{1}{2}\bp \cdot \mathsf{D} \bp.
\end{align}
The overdots signify the total time-derivative ${\rm d}/{\rm d} t$ along the characteristics $\bq(t)$.
In particular, $\dot{W}(\bq(t),t) = \partial W/\partial t + \dot{\bq} \cdot \partial W/\partial \bq$.
Equations \eqref{eq:qdot} and \eqref{eq:pdot} are Hamilton's equations, while Eq.~\eqref{eq:Wdot} is the differential equation satisfied by the classical action.

The Hamiltonian system Eq.~\eqref{eq:qdot} and \eqref{eq:pdot} stems from a variational principle, which sheds light on the physical meaning of the approximation \eqref{eq:pertexp}.
We define the action functional as the time integral of \eqref{eq:Wdot} for arbitrary functions of time $(\bq(\tau),\bp(\tau))$,
\begin{equation}\label{eq:action_princ1}
\mathcal{I}[\bq(\tau),\bp(\tau)] = \int_0^t \left[ \bp \cdot \dot{\bq} - H(\bq,\bp) \right] {\rm d} \tau.
\end{equation}
The critical points of this action functional are derived from the Euler-Lagrange equations of \eqref{eq:action_princ1}, which yield directly Eqs.~\eqref{eq:qdot} and \eqref{eq:pdot}.
Therefore, $W$ is simply the value of the action functional $\mathcal{I}$ evaluated at its critical points, up to an additive constant.
The phase-space action functional  \eqref{eq:action_princ1} can be converted into a configuration space action functional, by going from the Hamiltonian formulation to the Lagrangian formulation.
Assuming $\mathsf{D}$ is positive-definite and thus invertible, Eq.~\eqref{eq:qdot} can be solved for $\bp$, yielding $\bp = \mathsf{D}^{-1} [\dot{\bq} - \bF(\bq)]$.
Using the second equality of Eq.~\eqref{eq:Wdot}, we can now eliminate $\bp$ from Eq.~\eqref{eq:action_princ1}, yielding the configuration space action
\begin{equation}\label{eq:action_princ2}
\widetilde{\mathcal{I}}[\bq(\tau)] = \frac{1}{2}\int_0^t \left(\dot{\bq} - \bF(\bq)\right)\cdot \mathsf{D}^{-1}\left(\dot{\bq} - \bF(\bq)\right)   {\rm d} \tau.
\end{equation}
Equation \eqref{eq:action_princ2} is alternately known as the Onsager-Machlup action functional \cite{Onsager1953a} or the Freidlin-Wentzell action functional \cite{Freidlin2012}.
Because $\mathsf{D}$ is positive-definite, $\widetilde{\mathcal{I}} \geq 0$, with equality only achieved along the deterministic trajectories satisfying $\dot{\bq} = \bF(\bq)$.
Hence, $\widetilde{\mathcal{I}}$ is like a cost functional which penalizes deviations from the deterministic trajectories.
In the case of the propagator initial condition \eqref{eq:prop_ic}, the subsequent probability density \eqref{eq:propsemi1} is peaked along the deterministic trajectory, and deviations away from the trajectory due to noise are exponentially suppressed.
Furthermore, we can now see that the approximation implied by Eq.~\eqref{eq:propsemi1} is that only the critical points of the functional \eqref{eq:action_princ2} contribute to the probability density; all other paths $\bq(\tau)$ are discarded in this approximation.
This formulation also highlights the link with the path-integral formulation of the Fokker-Planck equation \cite{Langouche1978,Langouche1982}.

A useful concept for understanding the time evolution of the semiclassical probability density \eqref{eq:propsemi1} is the \emph{Lagrangian manifold} \cite{Littlejohn1992}.
According to Eqs.~\eqref{eq:qdot}--\eqref{eq:Wdot}, the action $W(\bq,t)$ is expressed in terms of families of solutions of a Hamiltonian system that, at any instant of time, lie on the $d$-dimensional surface in phase space defined by Eq.~\eqref{eq:lagman}.
Equation \eqref{eq:lagman} defines a surface which is a Lagrangian manifold, i.e.\ a surface in phase space on which the symplectic $2$-form vanishes (for more details, see \cite{Littlejohn1992}).
The time-evolution of $W$ (and $A$, as shown in Sec.~\ref{sec:A}) is thus directly obtained from the time evolution of a properly selected initial Lagrangian manifold  under the Hamiltonian flow.
The initial Lagrangian manifold, i.e.\ a $d$-dimensional surface in phase space containing the initial conditions $(\bq',\bp')$, must be selected so that the initial condition is satisfied, i.e.  $\lim_{t\rightarrow0} A(\bq,t)\exp[-W(\bq,t)/\varepsilon] = P_0(\bq)$.
We use primed variables to refer to the space of initial conditions.
Though Eq.~\eqref{eq:lagman} explicitly gives the relationship between $W$ and the Lagrangian manifold for times $t \neq 0$, it is ill-defined for any $P_0(\bq)$ for which $W$ is singular in the $t \rightarrow 0$ limit.
This occurs when the projection of the initial Lagrangian manifold into configuration space is singular, which is the case for the propagator and hybrid initial conditions.

For the propagator initial condition \eqref{eq:prop_ic}, the initial Lagrangian manifold is the phase-space surface $\bq' = \bq_0$.
This means the Lagrangian manifold includes all possible initial momenta $\bp' \in \mathbb{R}^d$.
The projection of this Lagrangian manifold into $\bq$ space is singular because all points of the Lagrangian manifold project to the same configuration space point, $\bq_0$.
The solution of Eq.~\eqref{eq:HJ} in this case is $W(\bq,t) = R(\bq,\bq_0,t)$, where $R$ is Hamilton's principal function, given by
\begin{equation}\label{eq:action}
R(\bq,\bq_0,t) = \int_0^t \left[ \bp(\tau) \cdot \dot{\bq}(\tau) - H(\bq(\tau),\bp(\tau)) \right] {\rm d} \tau,
\end{equation}
where $(\bq(\tau),\bp(\tau))$ is the Hamiltonian path [i.e.\ solution of Eqs.~\eqref{eq:qdot} and \eqref{eq:pdot}] going from $\bq_0$ to $\bq$ in time $t$.
Equation \eqref{eq:action} clearly follows from Eq.~\eqref{eq:Wdot}.
Because this solution is obtained by evolving the surface $\bq' = \bq_0$ forward in time, each path has a distinct initial momentum $\bp_0$.
In fact, $\bp_0$ may be expressed in terms of $\bq_0$, $\bq$, and $t$ as
\begin{equation}\label{eq:p0}
\bp_0 = -\frac{\partial R}{\partial \bq_0}(\bq,\bq_0,t).
\end{equation}
To prove it, we consider the change in $R$ as we make an infinitesimal change to the initial coordinate $\bq_0$, while keeping the final coordinate $\bq$ and transit time $t$ fixed.
We get
\begin{align*}
R(\bq,\bq_0 + \delta \bq_0,t) - R(\bq,\bq_0,t)  & \approx \int_0^t \left[ \delta \bp \cdot \dot{\bq} + \bp \cdot \delta  \dot{\bq} - \frac{\partial H}{\partial \bq} \cdot \delta \bq - \frac{\partial H}{\partial \bp} \cdot \delta \bp \right] {\rm d}\tau \\
& =  \bp \cdot \delta \bq \big|_0^t - \int_0^t \left[ \dot {\bp} \cdot \delta \bq + \frac{\partial H}{\partial \bq} \cdot \delta \bq \right] {\rm d} \tau \\
& = -\bp_0 \cdot \delta \bq_0.
\end{align*}
In Sec.~\ref{sec:norm}, we show that Eq.~\eqref{eq:action} indeed gives a probability density that satisfies the initial condition \eqref{eq:prop_ic}.

All of our arguments make the assumption that there is only one Hamiltonian path going from $\bq_0$ to $\bq$ in time $t$, so that $R$ is single-valued  (for all times $t > 0$).
In general, however, at longer times there are multiple paths connecting $\bq_0$ and $\bq$ in the same time $t$, with distinct initial momenta $\bp_0$.
Thus, the action becomes multi-valued.
This situation also arises in semiclassical quantum mechanics.
In that case, the quantum propagator consists of a sum of terms of the form of Eq.~\eqref{eq:propsemi1}, one for each branch of solutions, that are stitched together in such a way that the propagator is continuous \cite{Littlejohn1992,Maslov1981}.
We are not aware of a similar procedure for the Fokker-Planck equation.
Thus, we continue to assume that there is a unique characteristic for any $(\bq_0,\bq,t)$ and hence, $R$ is single-valued.

\subsection{Transport equation}\label{sec:A}
Next, we look at the equation arising from the first order terms of Eq.~\eqref{eq:consist}.
These lead to a transport equation for $S_1$, which can be rearranged into a transport equation for $A$.
This results in
\begin{equation}\label{eq:Atransport}
\dot{A} = -\left(\nabla \cdot \bF - \bG \cdot \nabla W+ \frac{1}{2}\mathsf{D} : \frac{\partial^2 W}{\partial \bq^2} \right) A,
\end{equation}
where we recall $\dot{A} = {\rm d} A/{\rm d} t = \partial A/\partial t + \dot{\bq} \cdot \nabla A$, with $\dot{\bq}$ given by Eq.~\eqref{eq:qdot}.
We solve Eq.~\eqref{eq:Atransport} by integrating along the characteristics $\bq(t)$ defined be Eqs.~\eqref{eq:qdot}--\eqref{eq:Wdot}.
Rearranging, we obtain
\begin{equation}\label{eq:density1}
\frac{{\rm d} (\ln A)}{{\rm d} t} = -\frac{1}{2} \nabla \cdot (\bF + \mathsf{D} \nabla W) -\frac{1}{2} \nabla \cdot \bF  + \bG \cdot \nabla W .
\end{equation}
Equation \eqref{eq:density1} gives the change of $A$ as one moves along a characteristic path from a point on the initial Lagrangian manifold, with configuration space coordinate $\bq'$, to the terminal coordinate $\bq$ in time $t$.
By integration, we obtain
\begin{equation}\label{eq:density2}
\ln\left(\frac{A(\bq,t)}{A(\bq',0)}\right) = -\frac{1}{2}\int_0^t \nabla \cdot (\bF + \mathsf{D} \nabla W) {\rm d} \tau - \frac{1}{2}\int_0^t \nabla \cdot \bF {\rm d}\tau + \int_0^t \bG \cdot \nabla W {\rm d} \tau,
\end{equation}
where the integration on the right-hand side of Eq.~\eqref{eq:density2} is carried out along the characteristic path.
Equation \eqref{eq:density2} applies to any initial Lagrangian manifold, but it must be handled carefully for initial Lagrangian manifolds with a singular projection, such as the propagator case for which $\bq' = \bq_0$.
The propagator initial condition \eqref{eq:prop_ic} is itself singular at $\bq' = \bq_0$, and it turns out that $A$ is also singular at this point.
We introduce the quantity $A_0 \equiv A(\bq_0,0)$ as a placeholder for now, and it is properly accounted for in Sec.~\ref{sec:norm}.

To clarify the first term on the right-hand side of Eq.~\eqref{eq:density2}, we introduce the phase space functions $(\bq(\bz',t),\bp(\bz',t))$.
The functions  $(\bq(\bz',t),\bp(\bz',t))$ express the positions and momenta $(\bq,\bp)$ at time $t$ of Eqs.~\eqref{eq:qdot} and \eqref{eq:pdot} as a function of their initial coordinate $\bz'$ on the Lagrangian manifold.
We allow for an arbitrary parametrization $\bz'$ of the initial Lagrangian manifold, and we express the initial conditions as $(\bq'(\bz'),\bp'(\bz'))$.
For the propagator initial condition, the initial Lagrangian manifold can be simply parametrized as 
\begin{equation}\label{eq:prop_param}
(\bq'(\bz'),\bp'(\bz')) = (\bq_0,\bz').
\end{equation}
The functions $(\bq,\bp)$ can be expressed using the flow functions $(\bQ(\bq',\bp',t),\bP(\bq',\bp',t))$, which map initial conditions $(\bq',\bp')$ to their values $\bQ$ and $\bP$ at time $t$ and satisfy Hamilton's equations \eqref{eq:qdot} and \eqref{eq:pdot}.
It is obvious that
\begin{subequations}\label{eq:flow_func}
\begin{align}\label{eq:qQ}
\bq(\bz',t) & = \bQ(\bq'(\bz'),\bp'(\bz'),t), \\ \label{eq:dpp}
\bp(\bz',t) & = \bP(\bq'(\bz'),\bp'(\bz'),t) = \nabla W (\bq(\bz',t),t).
\end{align}
\end{subequations}
From Eq.~\eqref{eq:flow_func}, it follows
\begin{subequations}\label{eq:dzdzp}
\begin{align}\label{eq:dqdp}
 \frac{\partial \bq}{\partial \bz'} & = \frac{\partial \bQ}{\partial \bq'} \frac{\partial \bq'}{\partial \bz'} +   \frac{\partial \bQ}{\partial \bp'} \frac{\partial \bp'}{\partial \bz'}, \\
\frac{\partial \bp}{\partial \bz'} & = \frac{\partial \bP}{\partial \bq'} \frac{\partial \bq'}{\partial \bz'} +   \frac{\partial \bP}{\partial \bp'} \frac{\partial \bp'}{\partial \bz'} .
\end{align}
\end{subequations}

Hence, we can compute the time evolution of the Jacobian matrix $\partial \bq/\partial \bz'$  by differentiating Eq.~\eqref{eq:dqdp} with respect to time.
This leads to 
\begin{align} 
\frac{{\rm d}}{{\rm d} t} \frac{\partial \bq}{\partial \bz'} & = \left(\frac{{\rm d}}{{\rm d} t} \frac{\partial \bQ}{\partial \bq'} \right) \frac{\partial \bq'}{\partial \bz'} +  \left( \frac{{\rm d}}{{\rm d} t} \frac{\partial \bQ}{\partial \bp'} \right) \frac{\partial \bp'}{\partial \bz'} \nonumber \\
& = \left( \frac{\partial \bF}{\partial \bq} \frac{\partial \bQ}{\partial \bq'} + \mathsf{D} \frac{\partial \bP}{\partial \bq'} \right) \frac{\partial \bq'}{\partial \bz'} + \left( \frac{\partial \bF}{\partial \bq} \frac{\partial \bQ}{\partial \bp'} + \mathsf{D} \frac{\partial \bP}{\partial \bp'} \right) \frac{\partial \bp'}{\partial \bz'} \nonumber \\ 
& = \frac{\partial \bF}{\partial \bq} \frac{\partial \bq}{\partial \bz'} + \mathsf{D} \frac{\partial \bp}{\partial \bz'} \nonumber \\ \label{eq:dqdpdot}
& = \left[\frac{\partial \bF}{\partial \bq} + \mathsf{D} \frac{\partial^2 W}{\partial \bq^2} \right]  \frac{\partial \bq}{\partial \bz'}.
\end{align}
We obtain the second line of Eq.~\eqref{eq:dqdpdot} by differentiating Eq.~\eqref{eq:qdot} with respect to $\bq'$ and $\bp'$ and the third line by applying Eq.~\eqref{eq:dzdzp}.
In the fourth line of Eq.~\eqref{eq:dqdpdot}, we use
\begin{equation}
\frac{\partial \bp}{\partial \bz'} = \frac{\partial^2 W}{\partial \bq^2} \frac{\partial \bq}{\partial \bz'},
\end{equation}
which follows from Eq.~\eqref{eq:dpp}.
Next, we find the equation satisfied by $D \equiv \det \partial \bq/\partial \bz'$, which is
\begin{equation}\label{eq:lndet}
\frac{{\rm d} (\ln |D|)}{{\rm d} t} = {\rm tr} \left[\left(\frac{{\rm d}}{{\rm d}t} \frac{\partial \bq}{\partial \bz'} \right) \left(\frac{\partial \bq}{\partial \bz'} \right)^{-1}\right] = \nabla \cdot ( \bF + \mathsf{D} \nabla W),
\end{equation}
where we have used Eq.~\eqref{eq:dqdpdot}.
From this it follows
\begin{equation}\label{eq:det}
\ln \left( \frac{|D|}{D_0} \right) = \int_0^t \nabla \cdot (\bF + \mathsf{D} \nabla W) {\rm d} \tau.
\end{equation}
The right-hand side of Eq.~\eqref{eq:det} is the first term that appears in Eq.~\eqref{eq:density2} up to a factor of $-1/2$.
In Eq.~\eqref{eq:det},
\begin{equation}
D_0 = \lim_{t \rightarrow 0} \left|\det \frac{\partial \bq}{\partial \bz'} \right|.
\end{equation}
For the case of the propagator initial condition, $\bq \rightarrow \bq_0$ as $t \rightarrow 0$, independent of $\bz'$,  and therefore $\partial \bq/\partial \bz' = 0$ in the limit, implying $D_0 = 0$.
This is related to the divergence of $A$ as $t \rightarrow 0$, so we keep $D_0$ as a placeholder here and return to this point in Sec.~\ref{sec:norm}.
Using Eq.~\eqref{eq:det}, we may now solve Eq.~\eqref{eq:density2} for $A$, giving
\begin{equation}\label{eq:amp1}
A(\bq,t) = A_0 \sqrt{ D_0}  \left| \det \frac{\partial \bq}{\partial \bp'} \right|^{-1/2} \exp\left[-\frac{1}{2} \int_0^t \nabla \cdot \bF {\rm d}\tau + \int_0^t \bG \cdot \bp\, {\rm d} \tau \right],
\end{equation}
where we have replaced $\bz'$ by $\bp'$ by virtue of Eq.~\eqref{eq:prop_param}.

\subsection{Normalization of the propagator}\label{sec:norm}
To fix the value of the constant $A_0\sqrt{D_0}$, we must consider the limit $t\rightarrow 0$ and impose the initial condition \eqref{eq:prop_ic}.
In this limit we have the estimates
\begin{align}\label{eq:qest}
\frac{\bq-\bq_0}{t} & \approx \bF(\bq_0) + \mathsf{D} \bp', \\
R & \approx \frac{1}{2t}(\bq - \bq_0) \cdot \mathsf D^{-1} (\bq - \bq_0).
\end{align}
Solving Eq.~\eqref{eq:qest} for $\bq$, we find
\begin{align}
\frac{\partial \bq}{\partial \bp'} & = \mathsf{D}t, \\
\det  \frac{\partial \bq}{\partial \bp'} & = t^d \det \mathsf{D} .
\end{align}
Hence, the semiclassical propagator in this limit becomes
\begin{equation}\label{eq:propnorm}
K(\bq,\bq_0,t\rightarrow 0) \approx A_0 \sqrt{D_0} \left(\det \mathsf{D} \right)^{-1/2} t^{-d/2} \exp\left[-\frac{1}{2\varepsilon t}(\bq - \bq_0) \cdot \mathsf{D}^{-1} (\bq - \bq_0)\right].
\end{equation}
Equation \eqref{eq:propnorm} is a Gaussian approximation to the $\delta$ function initial condition \eqref{eq:prop_ic}, which approaches the $\delta$ function in the $t\rightarrow 0$ limit.
Therefore, the solutions for $W$ and $A$ satisfy the initial condition, provided that
\begin{equation}\label{eq:normcoefficient}
A_0 \sqrt{ D_0} = (2\pi \varepsilon)^{-d/2},
\end{equation}
so that the Gaussian is properly normalized.
The final semiclassical expression for the propagator is
\begin{equation}\label{eq:final_prop}
K(\bq,\bq_0,t) \approx \frac{1}{(2\pi \varepsilon)^{d/2} } \left| \det \frac{\partial \bq}{\partial \bp'} \right|^{-1/2}\exp\left[-\frac{1}{\varepsilon}R(\bq,\bq_0,t) -\frac{1}{2}\int_0^t \nabla \cdot \bF {\rm d}\tau +  \int_0^t \bG \cdot \bp\, {\rm d}\tau\right].
\end{equation}

\subsection{Solutions for WKB and hybrid initial conditions}\label{sec:wkb}
\subsubsection{WKB initial condition}
For the WKB initial condition \eqref{eq:WKBic}, the initial Lagrangian manifold is the surface defined by
\begin{equation}\label{eq:lagmanWKB}
\bp' = \frac{\partial W_0}{\partial \bq'}(\bq').
\end{equation}
We parametrize the Lagrangian manifold by 
\begin{equation}\label{eq:WKB_param}
(\bq'(\bz'),\bp'(\bz')) = \left(\bz', \frac{\partial W_0}{\partial \bq'}(\bz')\right),
\end{equation}
and let the phase space functions be $(\bq(\bz',t),\bp(\bz',t))$.
The position coordinate part of this function is assumed to be invertible, so that the initial configuration space coordinate can be expressed as $\bq_0 = \bq_0(\bq,t)$.
Then, the solution to Eq.~\eqref{eq:HJ} is \cite{Littlejohn1992,ArnoldCM}
\begin{equation}\label{eq:actionWKB}
W(\bq,t) = W_0(\bq_0(\bq,t)) + R(\bq,\bq_0(\bq,t),t),
\end{equation}
where $R$ is evaluated along the Hamiltonian trajectory with initial coordinate $\bq_0(\bq,t)$ and initial momentum given by Eq.~\eqref{eq:lagmanWKB} evaluated at $\bq_0(\bq,t)$.

The solution to the transport equation \eqref{eq:Atransport} is almost identical to the propagator initial condition case.
The main differences are the parametrization of the Lagrangian manifold \eqref{eq:WKB_param} and the specific initial condition $A(\bq,0) = A_0(\bq)$.
Taking these into account, we obtain
\begin{equation}\label{eq:ampWKB}
A(\bq,t) = A_0(\bq_0(\bq,t))  \left| \det \frac{\partial \bq}{\partial \bq'} \right|^{-1/2} \exp\left[-\frac{1}{2} \int_0^t \nabla \cdot \bF {\rm d}\tau + \int_0^t \bG \cdot \bp\, {\rm d} \tau \right].
\end{equation}

The final expression for the semiclassical probability density is
\begin{equation}\label{eq:final_wkb}
P(\bq,t) \approx A_0(\bq_0)  \left| \det \frac{\partial \bq}{\partial \bq'} \right|^{-1/2} \exp\left[-\frac{(W_0(\bq_0) + R(\bq,\bq_0,t))}{\varepsilon} - \frac{1}{2}\int_0^t \nabla \cdot \bF {\rm d}\tau + \int_0^t \bG \cdot \bp\, {\rm d}\tau \right].
\end{equation}

\subsubsection{Hybrid initial condition}\label{sec:hybrid}
For the hybrid initial condition \eqref{eq:hybridic}, the initial Lagrangian manifold is the surface defined by
\begin{align}
\bq_f' & = \bq_{f0}, \\
\bp_k' & = \frac{\partial U}{\partial \bq_k}(\bq_k').
\end{align}
We parametrize the Lagrangian manifold by the coordinates $\bz' \equiv (\bp_f',\bq_k')$, and let the phase space functions be $(\bq(\bz',t),\bp(\bz',t))$.
We assume an inverse to $\bq(\bz',t)$ exists, which we denote $(\bp_{f0},\bq_{k0}) = \bz_0 = \bz_0(\bq,\bq_{f0},t)$.
Then, the solution to the Hamilton-Jacobi equation is
\begin{equation}
W(\bq,t) = U(\bq_{k0}) + R(\bq,\bq_0,t),
\end{equation}
where $\bq_0 = (\bq_{f0},\bq_{k0}(\bq,\bq_{f0},t))$, and the initial momentum of the trajectory terminating at $\bq$ at time $t$ is 
\begin{align}
\bp_{f0}(\bq,\bq_{f0},t) &= -\frac{\partial R}{\partial \bq_{f0}}(\bq,\bq_0,t), \\
\bp_{k0}(\bq,\bq_{f0},t) & = \frac{\partial U}{\partial \bq_k}(\bq_{k0}).
\end{align}

Solving the transport equation  is again similar to the propagator case.
The quantity $\det \partial \bq/\partial \bz'$ still satisfies Eq.~\eqref{eq:lndet}, while in Eq.~\eqref{eq:density2} we write $A(\bq',0) = A_0(\bq_k')A_*$, where $A_*$ is a placeholder constant.
We therefore obtain
\begin{equation}\label{eq:A_hybrid}
A(\bq,t) = A_* \sqrt{D_0} A_0(\bq_{k0}(\bq,\bq_{f0},t))\left|\det \frac{\partial \bq}{\partial \bz'}\right|^{-1/2} \exp \left[-\frac{1}{2} \int \nabla \cdot \bF {\rm d} \tau + \int_0^t \bG \cdot \bp {\rm d}\tau \right].
\end{equation}
Taking the $t \rightarrow 0$ limit of the full solution and forcing it to satisfy the initial condition \eqref{eq:hybridic} leads to
\begin{equation}
A_*\sqrt{ D_0} = (2 \pi \varepsilon)^{-d_f/2}.
\end{equation}
Thus, the full semiclassical probability density is
\begin{align}
P(\bq,t) = & \frac{1}{(2\pi\varepsilon)^{d_f/2}} A_0(\bq_{k0})\left|\det \frac{\partial \bq}{\partial \bz'}\right|^{-1/2} \times \nonumber \\ \label{eq:final_hybrid}
&  \exp \left[-\frac{(U(\bq_{k0}) + R(\bq,\bq_0,t))}{\varepsilon}-\frac{1}{2} \int \nabla \cdot \bF {\rm d} \tau + \int_0^t \bG \cdot \bp {\rm d}\tau \right].
\end{align}

\subsection{Gaussian approximation to the semiclassical propagator: 1D case}\label{sec:gauss_approx}
For sufficiently small noise and times, it is useful to approximate the semiclassical propagator as a Gaussian centered on the deterministic trajectory $\bq^*(t)$.
Because of the exponential form of Eq.~\eqref{eq:final_prop}, for short times and small noise, the most important part to capture is the part near the absolute minimum of the action $R(\bq,\bq_0,t) = 0$, which due to Eq.~\eqref{eq:action_princ2}, occurs at the deterministic solution $\bq = \bq^*(t)$.
We illustrate the approximation for the 1D case for simplicity, so that $\bq \rightarrow q$ and $\bF \rightarrow F$, and we let the diffusion tensor $\mathsf{D} \rightarrow 1$ and $\bG \rightarrow 0$.
Thus, Hamilton's equations become
\begin{align}\label{eq:ham1dq}
\dot{q} = F + p, \\ \label{eq:ham1dp}
\dot{p} = -p \frac{dF}{dq}.
\end{align}
We expand $K$ about $q^*$ as follows:
\begin{equation}\label{eq:gauss_approx}
K(q,q_0,t) \approx A(q^*) \exp\left[ - \frac{1}{2\varepsilon} \frac{\partial^2 R}{\partial q^2} (q - q^*)^2\right],
\end{equation}
where
\begin{equation} \label{eq:gauss_approx_A}
A(q^*) = \frac{1}{\sqrt{2\pi\varepsilon}} \left| \frac{\partial q}{\partial p'} \right|^{-1/2} \exp\left[ -\frac{1}{2}\int_0^t \frac{d F}{d q} {\rm d} \tau \right].
\end{equation} 
In Eq.~\eqref{eq:gauss_approx_A}, the Jacobian matrix and integral are evaluated along the deterministic trajectory $q^*(t)$, so that $A(q^*)$ is a function of time.
Equation \eqref{eq:gauss_approx} thus constitutes a Gaussian approximation to the propagator, which would be properly normalized provided that
\begin{equation}\label{eq:gauss_norm_cond}
\sqrt{\frac{\partial^2 R}{\partial q^2}} = \left| \frac{\partial q}{\partial p'} \right|^{-1/2} \exp\left[ -\frac{1}{2}\int_0^t \frac{d F}{d q} {\rm d} \tau \right].
\end{equation}
Next, we show that this is indeed the case.

Recalling that $p = \partial R/\partial q$, we have that
\begin{equation}\label{eq:d2Rdq2}
\frac{\partial^2 R}{\partial q^2} = \frac{\partial p}{\partial q}.
\end{equation}
We now rewrite Eq.~\eqref{eq:d2Rdq2} in terms of partial derivatives with respect to the initial conditions $(q',p')$, which we can then compute using the tangent flow equations of Eqs.~\eqref{eq:ham1dq} and \eqref{eq:ham1dp}.
Using the chain rule, we obtain
\begin{equation}\label{eq:dpdq}
\frac{\partial p}{\partial q} = \frac{\partial p}{\partial p'} \left( \frac{\partial q}{\partial p'} \right)^{-1}
\end{equation}
The quantity $\partial p/\partial p' $ satisfies
\begin{equation}\label{eq:dpdp'dt}
\frac{\rm d}{{\rm d} t} \frac{\partial p}{\partial p'} = -p \frac{d^2 F}{d q^2}\frac{\partial q}{\partial p'} - \frac{d F}{d q} \frac{\partial p}{\partial p'},
\end{equation}
with initial condition $\partial p/\partial p'(0) = 1$.
Along the deterministic trajectory however, $p = 0$ for all time, so the first term of Eq.~\eqref{eq:dpdp'dt} vanishes.
Therefore, we obtain
\begin{equation}\label{eq:dpdp'}
\frac{\partial p}{\partial p'} = \exp\left[ -\int_0^t \frac{d F}{d q} {\rm d} \tau \right].
\end{equation}
Combining Eqs.~\eqref{eq:d2Rdq2}, \eqref{eq:dpdq}, and \eqref{eq:dpdp'}, we see that Eq.~\eqref{eq:gauss_norm_cond} is almost proved.
We need only verify that $\partial q/\partial p' \geq 0$ for all time.
This quantity satisfies
\begin{equation}
\frac{\rm d}{{\rm d}t} \frac{\partial q}{\partial p'} = \frac{d F}{d q} \frac{\partial q}{\partial p'} + \frac{\partial p}{\partial p'},
\end{equation}
with initial condition $\partial q/\partial p'(0) = 0$.
Using Eq.~\eqref{eq:dpdp'}, we obtain
\begin{equation}\label{eq:dqdp'}
\frac{\partial q}{\partial p'} = \exp \left[ \int_0^t \frac{dF}{dq} {\rm d} \tau \right] \int_0^t \exp\left[ - 2 \int_0^{\tau'} \frac{dF}{dq} {\rm d} \tau''\right] {\rm d} \tau'.
\end{equation}
Because Eq.~\eqref{eq:dqdp'} consists of products and sums of exponentials, which are all positive, we have $\partial q/\partial p' \geq 0$. Thus, Eq.~\eqref{eq:gauss_norm_cond} is proved.

Combining the results, we obtain the following expression for the Gaussian approximation to the propagator:
\begin{equation}\label{eq:gauss_prop}
K(q,q_0,t) = \sqrt{\frac{1}{2\pi \varepsilon} \frac{\partial^2 R}{\partial q^2}}\exp\left[ - \frac{1}{2\varepsilon} \frac{\partial^2 R}{\partial q^2} (q - q^*)^2\right], 
\end{equation}
where
\begin{equation}\label{eq:gauss_var}
\frac{\partial^2 R}{\partial q^2} = \exp\left[ -2\int_0^t \frac{d F}{d q} {\rm d} \tau \right] \left\{\int_0^t \exp\left[ - 2 \int_0^{\tau'} \frac{dF}{dq} {\rm d} \tau''\right] {\rm d} \tau' \right\}^{-1}.
\end{equation}

\section{Solution to the Liouville equation}\label{sec:liou_solu}
We derive the solution to Eq.~\eqref{eq:theta_liou}, rewritten here as
\begin{equation}\label{eq:liou_app}
\frac{\partial P}{\partial \tau} -\alpha \sin (2\theta) \frac{\partial P}{\partial \theta} = 2 \alpha \cos (2 \theta) P.
\end{equation}
The method of characteristics seeks a solution $P(\theta(\tau),\tau)$, where the characteristics $\theta(\tau)$  satisfy
\begin{equation}
\frac{{\rm d} \theta}{{\rm d} \tau} = -\alpha \sin (2\theta)
\end{equation}
and are explicitly given by Eq.~\eqref{eq:theta_t_exact}.
Taking the total time derivative of  $P(\theta(\tau),\tau)$ and using Eq.~\eqref{eq:liou_app}, we get
\begin{equation}\label{eq:liou_charac}
\frac{{\rm d} P}{\rm d \tau} = 2 \alpha \cos (2 \theta(\tau)) P.
\end{equation}
We use the identity $\cos 2\theta = (1-\tan^2\theta)/(1+\tan^2\theta)$, substitute $\tan(\theta(\tau)) = e^{-2\alpha \tau} \tan \theta_0$ [from Eq.~\eqref{eq:theta_t_exact}] and move $P$ to the left-hand side of Eq.~\eqref{eq:liou_charac}, which yields
\begin{equation}
\frac{{\rm d} (\ln P)}{\rm d \tau} = 2 \alpha \frac{1 - e^{-4 \alpha \tau} \tan^2 \theta_0}{1 + e^{-4 \alpha \tau} \tan^2 \theta_0}.
\end{equation}
Integrating both sides yields
\begin{equation}\label{eq:liou_pen}
\ln \frac{P(\theta,\tau)}{P_0(\theta_0)} = \frac{1}{2} \ln \left[ \frac{e^{4\alpha \tau} + 2 \tan^2 \theta_0  + e^{-4 \alpha \tau} \tan^4 \theta_0}{\left( 1 + \tan^2 \theta_0\right)^2} \right].
\end{equation}
Finally, we solve Eq.~\eqref{eq:theta_t_exact} for $\tan \theta_0$ and $\theta_0$ in terms of $\theta$ and $\tau$, substitute these into Eq.~\eqref{eq:liou_pen}, and solve for $P(\theta,\tau)$, which yields
\begin{equation}
P(\theta,\tau) = P_0\left(\tan^{-1}\left(e^{2  \alpha \tau} \tan \theta \right) \right)\frac{e^{2\alpha\tau}}{\cos^2 \theta +e^{4 \alpha \tau}\sin^2\theta }.
\end{equation}


\end{document}